\documentclass[12pt]{article}
\usepackage{amsfonts}
\usepackage{color}
\usepackage{graphicx}
\usepackage{amsmath}
\usepackage{float}
\usepackage{amssymb}
\providecommand{\U}[1]{\protect\rule{.1in}{.1in}}
\textwidth=6.5in \hoffset=-.75in \voffset=-1in \numberwithin{equation}{section}
\numberwithin{figure}{section}

\newcommand {\be}{\begin{equation}}
 \newcommand {\ee}{\end{equation}}
 \newcommand {\bea}{\begin{eqnarray}}
 
 \newcommand {\eea}{\end{eqnarray}}

 \newcommand {\nn}{\nonumber}

\begin{document}

\begin{titlepage}
\bigskip \begin{flushright}
\end{flushright}
\vspace{1cm}
\begin{center}
{\Large \bf {Gravitational Perturbation and Kerr/CFT Correspondence}}\\
\vskip 1cm
\end{center}
\vspace{1cm}
\begin{center}
A. M.
Ghezelbash{ \footnote{amg142@campus.usask.ca}}
\\
Department of Physics and Engineering Physics, \\ University of Saskatchewan,
Saskatoon, Saskatchewan S7N 5E2, Canada\\
\vspace{1 cm}
\end{center}
\begin{abstract}

We find the explicit form of two-point function for the conformal spin-2 energy momentum operators on the near horizon of a near extremal Kerr black hole by variation of a proper boundary action. In this regard, we consider an appropriate boundary action for the gravitational perturbation of the Kerr black hole. We show that the variation of the boundary action with respect to the boundary fields yields the two-point function for the energy momentum tensor of a conformal field theory. We find agreement between the two-point function and the correlators of the dual conformal field theory to the Kerr black hole. 
\end{abstract}
\bigskip
\end{titlepage}\onecolumn

\bigskip

\section{Introduction}

The holography for rotating black holes provides a description of black hole physics in terms of a certain two-dimensional conformal field theory (CFT). The first realization of the holography for rotating black holes was proposed for extremal Kerr black hole in \cite{Stro1}. The proposal states that the near horizon geometry of the extremal Kerr black hole does have a dual holographic description in terms of a two-dimensional chiral CFT. 
The proposal has been studied and extended extensively for other extremal or near extremal rotating black holes \cite{GH0}-\cite{Iso4}. For all extremal and near-extremal black holes in different dimensions, the physical quantities associated to black hole (such as the entropy or scattering cross section) are in agreement with the corresponding microscopic dual quantities in CFT. 
Moreover, inspired by the fact that near horizon region of a near extremal Kerr black hole contains a two dimensional copy of Anti de Sitter, a field theoretic calculation can be done to find the 
two-point correlation functions of conformal fields with different spin. Explicitly, in \cite{Becker:2012vda}, the two-point correlation functions of spinor fields were constructed by calculating the on-shell gravitational action for the spinor fields with appropriate boundary condition. Though the on-shell bulk action for the spinor fields vanishes, however one can add a boundary term to the gravitational action which doesn't vanish on-shell \cite{Hen}. The variation of the boundary action yields the proper two-point correlation functions for the conformal operators. In a recent article \cite{spin1}, the authors used an appropriate boundary action for the vector fields to calculate the two-point functions of conformal vector fields. In this article, we derive the two-point function for the energy-momentum tensor fields in the background of Kerr black hole. The first step is to find an appropriate boundary action for the spin-2 fields. In this regard, we write the Einstein-Hilbert action as the summation of two actions. The first action is a bulk action that unlike the Einstein-Hilbert action depends only on the metric tensor and its first derivative. Although the bulk action is not a scalar under the general coordinate transformations, however it leads to the same Einstein's field equations. The second action is a boundary action that depends on the metric tensor and the Christofel symbols. We consider the boundary action and calculate it on-shell to find the two-point functions for the spin-2 energy momentum tensor fields, on the near horizon region of the Kerr black hole. We then compare the result of our calculation to the correlation functions for spin-2 conformal operators in a conformal field theory. 

The paper is organized as follows. In section \ref{sec-BA}, we consider an appropriate boundary action to calculate the two-point functions of conformal energy-momentum tensor operators. In section \ref{sec-GP}, we review the gravitational perturbation of the Kerr black hole in Newman-Pernrose formalism and present the radial and angular Teukolsky equations. We notice that two Weyl scalars have separable decompositions in terms of radial and angular coordinates. We also present the explicit expressions for the spin coefficients in terms of the Teukolsky functions and their derivatives. We then calculate and present explicitly all the components of the gravitational perturbation field, as a bilinear form in terms of the radial and angular Teukolsky functions. Moreover, as the boundary action is over the horizon of the black hole, we present approximate solutions for the radial Teukolsky functions. In section \ref{sec-bacalc}, we explicitly calculate the boundary action on-shell and show that taking the functional derivatives of the action with respect to the boundary fields (that are dual to the conformal energy-momentum tensor operators), yields the two-point function of boundary conformal operators. We discuss about the results and note that one part of the two-point function can be written in terms of the retarded Green's function for the spin-2 fields, in agreement with previous suggestions in \cite{Hartman:2009nz}. In section \ref{sec-CFT}, we show that our field-theoretical two-point function can be matched to the Fourier transform of the finite temperature two-point function for spin-2 operators, by proper choice of conformal weights.

\section{Boundary action for gravitational perturbations of a black hole }
\label{sec-BA}
The gravitational action for a general four-dimensional curved spacetime ${\cal M}$ with a boundary $\partial M$ is given by 
\be
S=S_{EH}+S_{GH},\label{Stot}
\ee
where $S_{EH}=\frac{1}{16\pi}\int_{\cal M} d^4 x \sqrt{-g}R$ and $S_{GH}$ is the Gibbons-Hawking boundary action. The bulk action $S_{EH}$ depends on the second order derivatives of the metric field $g_{\mu\nu}$. However, we can use the following identity (\ref{SS1}) to obtain a bulk action that depends only on the metric field and the first order derivatives of the metric field. In fact,
\be
S_{EH}= S_1+\frac{1}{16\pi}\int_{\partial M} d^3x \sqrt{-g}\{g^{\mu\nu}\Gamma^\sigma _{\nu\sigma}-g^{\sigma\nu}\Gamma^\mu _{\sigma\nu}\},\label{SS1}
\ee
where $S_1=\frac{1}{16\pi}\int_{\cal M} d^4x \sqrt{-g} g^{\mu\nu}\{ \Gamma ^\tau_{\mu\nu}\Gamma ^\sigma_{\tau\sigma}-\Gamma ^\tau_{\mu\sigma}\Gamma ^\sigma_{\tau\nu}\}$.
The bulk action $S_1$ clearly depends on the gravitational field and its first derivatives and resembles to the action for a scalar field or spinor or electromagnetic field that only depends on dynamical field and its first derivatives. We note that the Einstein-Hilbert actions $S_{EH}$ and $S_1$ (together with the Gibbons-Hawking boundary action) lead to the same Einstein field equations for the gravitational field, however
the Einstein-Hilbert action $S_{EH}$ is definitely invariant under the general coordinate transformations, while $S_1$, in general is not invariant. In this aricle, we consider the action (\ref{Stot}) in the background of the Kerr black hole and use the boundary part of the action (\ref{SS1}) 
\be
S_B=\frac{1}{16\pi}\int_{r=r_B} d^3x \sqrt{-g}\{g^{r\nu}\Gamma^\sigma _{\nu\sigma}-g^{\sigma\nu}\Gamma^r _{\sigma\nu}\},\label{boundact}
\ee
to calculate the two point functions of spin 2 gravitational perturbation fields $\tilde h_{\mu\nu}$, where $g_{\mu\nu}=g^{(0)}_{\mu\nu}+\epsilon \tilde h_{\mu\nu}$. In (\ref{boundact}), the hypersurface $r=r_B$ is the boundary of near-NHEK geometry of Kerr black hole. The $g^{(0)}_{\mu\nu}$ describes the Kerr black hole which the line element in Boyer-Lindquist coordinates is given by
\be
ds^2 = - \frac{\Delta }
{{\rho ^2 }}\left( {dt - a\sin ^2 \theta d\phi } \right)^2 + \frac{{\rho ^2 }}
{\Delta }dr^2 + \rho ^2 d\theta ^2 + \frac{{\sin ^2 \theta }}
{{\rho ^2 }}\left( {adt - \left( {r^2 + a^2 } \right)d\phi } \right)^2 \label{Kerr-BL},
\ee
where $\rho^2 =\bar \rho \bar \rho ^*, \, \bar \rho= r + ia \cos \theta $ and $\Delta = r^2 + a^2 - 2Mr$. Moreover the field $\tilde h_{\mu\nu}$ is the spin 2 field which represents the gravitational perturbation of the Kerr black hole (\ref{Kerr-BL}). We present explicitly the different components of $\tilde h_{\mu\nu}$ in section \ref{sec-GP}. 

The Newman-Penrose null vectors $l^i,i=1,\cdots,4$ for the Kerr black hole (\ref{Kerr-BL}) are given by %
\begin{eqnarray}
l^1 &=& \Delta ^{ - 1} \left( {r^2 + a^2 ,\Delta ,0,a} \right),\label{l1}\\
l^2 &=& \frac{1}{{2\rho ^2 }}\left( {r^2 + a^2 , - \Delta ,0,a} \right),\label{l2}\\
l^3 &=& \frac{1}{{\bar \rho \sqrt 2 }}\left( {ia\sin \theta ,0,1,\frac{i}{{\sin \theta }}} \right),\label{l3}\\
l^4 &=& \frac{1}{{\bar \rho ^*\sqrt 2 }}\left( {-ia\sin \theta ,0,1,-\frac{i}{{\sin \theta }}} \right).\label{l4}
\end{eqnarray}
In Newman-Penrose null-basis (\ref{l1})-(\ref{l4}), the Weyl scalars $\Psi_0$,$\Psi_1$,$\Psi_3$ and $\Psi_4$ and spin coefficients $\kappa, \sigma, \lambda$ and $\nu$ for the Kerr black hole (\ref{Kerr-BL}) are zero, while $\Psi_2$ and the other eight spin-coefficients are not zero (Appendix \ref{Ap1}).
However, if we consider the gravitational perturbation of the Kerr black hole (\ref{Kerr-BL}) by a field of spin 2, the Weyl scalars $\Psi_0$,$\Psi_1$,$\Psi_3$ and $\Psi_4$ get first-order corrections. Moreover, 
the Weyl scalar $\Psi_2$ and all the spin coefficients get first-order corrections. We show the first-order corrections to any physical quantity by a superscript $(1)$, for example $\Psi_0^{(1)}$. 
The Kerr black hole (\ref{Kerr-BL}) is stationary and axisymmetric and so, we consider the dependence of a spin 2 field to $t$ and $\phi$ coordinates, in the background of the Kerr black hole, as 
\be {\bf h}_{\mu \nu} (t,r,\theta,\phi)=h_{\mu\nu}(r,\theta)e^{ - i\omega t + im\phi },\label{hbold}
\ee 
where $m = 0,1,2,...$ is the azimuthal number. We consider the different components of the gravitational perturbations $\tilde h_{\mu\nu}$ as the real parts of ${\bold h}_{\mu\nu}(t,r,\theta,\phi)$. 
In the Newman-Penrose null basis (\ref{l1})-(\ref{l4}), the components $h^{\mu\nu}(r,\theta)$ are given by \cite{Chan}
\be
h^{\mu \nu}=l^{\mu}n^{\nu(1)} +l^{\mu(1)}n^{\nu}+
l^{\nu(1)}n^{\mu}+l^{\nu}n^{\mu(1)}-(m^{\mu}{\bar m}^{\nu(1)}+{ m}^{\nu}{\bar m}^{\mu(1)}+m^{\nu(1)}{\bar m}^{\mu}+m^{\mu(1)}{\bar m}^{\nu}),\label{hmunu}
\ee
\nopagebreak
where for simplicity in notation, we denote the null vectors $l^1,\cdots ,l^4$ by $l,n,m,{\bar m}$ respectively.
The vectors $l^{\mu(1)}$ are, in general, a linear combination of the Newman-Penrose null-basis vectors $l^{\nu}$. We show the transformation matrix by ${\bold A}$ that transforms the null vector basis $l^i,i=1,\cdots ,4$ into the null basis for the perturbation $l^{\mu(1)}$ as $l^{\mu(1)}=A_\nu^\mu l^\nu$. 
\nopagebreak

\section{Gravitational perturbation field}
\label{sec-GP}

To evaluate the on-shell boundary action (\ref{boundact}), we use the explicit representation 
for the gravitational perturbation (\ref{hmunu}). The different components of $h^{\mu\nu}$ are given by \cite{Chan}
\begin{eqnarray}
h^{tt}&=&\frac{(r^2+a^2)^2}{\rho^2\Delta}(A_1^1+A_2^2+F_2^1+F_1^2)-\frac{a^2\sin^2\theta}{\rho^2}(A_3^3+A_4^4-\rho^2(F_4^3+F_3^4))
\nn\\&+&\frac{\sqrt{2}}{\Delta}ia(r^2+a^2)(C_1+B_2)\sin^2\theta,\label{htt1}\\
h^{rr}&=&-\frac{\Delta}{\rho^2}(A_1^1+A_2^2-F_1^2-F_2^1),\label{hrr1}\\
h^{\theta\theta}&=&-\frac{1}{\rho^2}(A_3^3+A_4^4)-F_4^3+F_3^4,\label{hthth1}\\
h^{\phi\phi}&=&\frac{a^2}{\rho^2\Delta}(A_1^1+A_2^2+F_1^2+F_2^1)-\frac{1}{\rho^2\sin^2\theta}(A_3^3+A_4^4-\rho^2(F_3^4+F_4^3))
+\frac{\sqrt{2}}{\Delta\sin\theta}ia(C_1+B_2),\nn \\&&\label{hpp1}\\
h^{tr}&=&-\frac{r^2+a^2}{\rho^2}(F_2^1-F_1^2)+\frac{ia\sin\theta}{\sqrt{2}}(H-J-C_2)\label{htr1},\\
h^{t\theta}&=&ia(F_4^3-F_3^4)\sin\theta+\frac{r^2+a^2}{\sqrt{2}\Delta}(F+G-B_1),\label{htth1}\\
h^{t\phi}&=&\frac{(r^2+a^2)a}{\rho^2\Delta}(A_1^1+A_2^2+F_1^2+F_2^1)-\frac{a}{\rho^2}(A_3^3+A_4^4-\rho^2F_3^4-\rho^2F_4^3)
\nn\\
&+&\frac{i}{\sqrt{2}\Delta\sin\theta}\{r^2+a^2(1+\sin^2\theta)\}(C_1+B_2),\label{htp1}\\
h^{r\theta}&=&-\frac{1}{\sqrt{2}}(F-G+H+J),\label{hrth1}\\
h^{r\phi}&=&-\frac{a}{\rho^2}(F_2^1-F_1^2)+\frac{i}{\sqrt{2}\sin\theta}(H-J-C_2),\label{hrp1}\\
h^{\theta\phi}&=&\frac{i(F_4^3-F_3^4)}{\sin\theta}+\frac{a}{\sqrt{2}\Delta}(F+G-B_1).\label{hthp1}
\end{eqnarray}
In equations (\ref{htt1})-(\ref{hthp1}), the different components of the matrix ${\bold F}$ 
are related to the components of ${\bold A}$, up to different multiplicative functions. In fact, we have
\be
F_1^2=\frac{2\rho^2}{\Delta}A_1^2, F_2^1=\frac{\Delta}{2\rho^2}A_2^1,F_3^4=\frac{1}{{\bar \rho} ^2}A_3^4,F_4^3==\frac{1}{({\bar \rho}^*) ^2}A_3^4.
\ee
The other quantities $F,G,H,J,B_1,B_2,C_1,C_2$ in equations (\ref{htt1})-(\ref{hthp1}), are also certain linear combinations of the different components of ${\bold A}$. As an example, $B_1=F_1^3+F_2^3+F_1^4+F_2^4$.
The Weyl scalars $\Psi_0^{(1)}$, $\Psi_1^{(1)}$ and the spin connections $\kappa^{(1)},\sigma^{(1)}$ satisfy three coupled linear Newman-Penrose differential equations. Similarly the other two Weyl scalars $\Psi_3^{(1)}$, $\Psi_4^{(1)}$ and the spin connections $\lambda^{(1)},\nu^{(1)}$ satisfy three coupled linear Newman-Penrose differential equations. 
The most important result is that one can decouple the set of first three differential equations as well as the second set of differential equations. In fact, we get two decoupled differential equations for the Weyl scalars $\Psi_0^{(1)}$ and $\Psi_4^{(1)}$ as
\begin{align}
\left( {\Delta \mathcal{D}_1 \mathcal{D}^\dag _2 + \mathcal{L}^\dag _{-1} \mathcal{L}_2 -6i\omega \bar \rho} \right)\Psi_0^{(1)} = 0,\label{PSI0}
\end{align}
and
\begin{align}
\left( {\Delta \mathcal{D}^\dag _{-1} \mathcal{D}_0 + \mathcal{L}_{-1} \mathcal{L}^\dag _2 +6 i\omega \bar \rho} \right)(\bar \rho^*)^4\Psi _4^{(1)} = 0.\label{PSI4}
\end{align}
In equations (\ref{PSI0}) and (\ref{PSI4}), the differential operators $\mathcal{D}_n, \mathcal{D}_n^\dag ,\mathcal{L}_n, \mathcal{L}_n^\dag $ where $n \in \mathbb{Z}$, are given by
\begin{equation}
\mathcal{D}_n = \frac{\partial }{{\partial r}} + \frac{{iK}}{\Delta } + 2n\left( {\frac{{r - M}}{\Delta }} \right),\, \mathcal{D}_n^\dag = \frac{\partial }{{\partial r}} - \frac{{iK}}{\Delta } + 2n\left( {\frac{{r - M}}{\Delta }} \right),
\label{DDdagger1}
\end{equation}
\begin{equation}
\mathcal{L}_n = \frac{\partial }{{\partial \theta }} + Q + n\cot \theta ,\,\mathcal{L}_n^\dag = \frac{\partial }{{\partial \theta }} - Q + n\cot \theta.
\label{LLdagger1}
\end{equation}
In equations (\ref{DDdagger1}) and (\ref{LLdagger1}), the $r$-dependent function $K$ is given by $K = - \left( {r^2 + a^2 } \right)\omega + am$ and the $\theta$-dependent function $Q$ is given by $ Q = - a\omega \sin \theta + \frac{m}{ {\sin \theta }}$. 
We can separate the variables in the partial differential equations (\ref{PSI0}) and (\ref{PSI4}) according to
\be
\Psi_0^{(1)}=R_{+2}(r)S_{+2}(\theta),\label{si0}
\ee
and
\be
\Psi_4^{(1)}=\frac{1}{(\bar \rho ^*)^4}R_{-2}(r)S_{-2}(\theta),\label{si4}
\ee
where $R_{\pm 2}$ and $S_{\pm 2}$ are called the Teukolsky functions. After substituting (\ref{si0}) and (\ref{si4}) in (\ref{DDdagger1}) and (\ref{LLdagger1}), respectively, we find that the radial Teukolsky functions $R_{\pm 2}$ satisfy the equations
\begin{eqnarray}
\left( {\Delta \mathcal{D}_1 \mathcal{D}^\dag _2 -6i\omega r } \right)R_{+2} &=& \bar \lambda R_{+2},\label{TeuR1}\\
\left( {\Delta \mathcal{D}^\dag _{-1}\mathcal{D}_0 +6i\omega r } \right)R_{-2} &=& \bar \lambda R_{-2},
\label{TeuR2}
\end{eqnarray}
and the angular Teukolsky functions $S_{\pm 2}$ satisfy the equations
\begin{eqnarray}
\left( {\mathcal{L}^\dag _{-1}\mathcal{L}_2 + 6a\omega \cos \theta } \right)S_{+2} &=& - \bar \lambda S_{+2},\\
\left( {\mathcal{L}_{-1} \mathcal{L}^\dag _2 -6a\omega \cos \theta } \right)S_{-2} &=& - \bar \lambda S_{-2}.
\label{TeuS2}
\end{eqnarray}
We note that in equations (\ref{TeuR1})-(\ref{TeuS2}), $\bar \lambda$ denotes the separation constant.
We can utilize the ten degrees of freedom that exist in describing the gravitational perturbation to fix the gauge freedom and so set some conditions on physical quantities. These ten degrees of freedom include six degrees of freedom corresponding to the rotation of the tetrad basis and four degrees of freedom corresponding to the translation of the coordinates. A straightforward and lengthy calculation shows that the Weyl scalars $\Psi_0^{(1)}$ and $\Psi_4^{(1)}$ 
are gauge invariant, however $\Psi_1^{(1)}$ and $\Psi_3^{(1)}$ are not gauge invariant. So, we use four degrees of freedom corresponding to the rotations of tetrad basis, to set $\Psi_1^{(1)}=\Psi_3^{(1)}=0$. This is a a proper choice for the Weyl scalars. A great advantage of choosing this gauge is that the spin coefficients can be written directly in terms of the radial and angular Teukolsky functions. In fact, we find the following relations for spin coefficients $\kappa^{(1)},\sigma^{(1)},\lambda^{(1)}$ and $\nu^{(1)}$ in terms of the Teukolsky functions and their derivatives,
\be
\kappa^{(1)}=-\frac{\sqrt{2}}{6M}(\bar\rho^*)^2 R_{+2}({\cal{L}}_2-\frac{3ia\sin\theta}{\bar \rho ^*})S_{+2},\label{sc1}
\ee
\be
\sigma^{(1)}=\frac{1}{6M\bar \rho}(\bar\rho^*)^2 S_{+2}\Delta({\cal{D}}^\dag _2-\frac{3}{\bar \rho ^*})R_{+2},\label{sc2}
\ee
\be
\lambda^{(1)}=\frac{1}{3M\bar \rho^*} S_{-2}({\cal{D}} _0-\frac{3}{\bar \rho ^*})R_{-2},\label{sc3}
\ee
\be
\nu^{(1)}=\frac{\sqrt{2}}{6M\rho ^2} R_{-2}({\cal{L}}^\dag _2-\frac{3ia\sin\theta}{\bar \rho ^*})S_{-2}.\label{sc4}
\ee
Moreover, we can use two coordinate degrees of freedom to set $\Psi_2^{(1)}=0$. We then have two remaining tetrad degrees of freedom and two coordinate degrees of freedom. We can use these four degrees of freedom to set
all the diagonal elements of matrix $\bf A$ equal to zero; $A_i^i=0,i=1,\cdots, 4$.
The function $B_2$ in equations (\ref{htt1}), (\ref{hpp1}) and (\ref{htp1}) is given by \cite{Chan}
\be
B_2=\frac{\Delta}{iK}\{\frac{a(K+\rho^2\omega)\cos\theta}{r\rho^2KQ\sin\theta}Z_2+\frac{1}{K\cos\theta}\{X_2-(\frac{r\omega}{K}+\frac{r}{\rho^2}-\frac{r-M}{\Delta})Z_1\}\},
\ee
in terms of three new functions $Z_1(r,\theta)=K(J-H)\cos\theta,Z_2(r,\theta)=-irQ(F+G)\sin\theta$ and $X_2(r,\theta)$. In what follows, we notice that these new functions can be expresses in terms of Teukolsky functions. 
Moreover, function $C_1$ in equations (\ref{htt1}), (\ref{hpp1}) and (\ref{htp1}) is given by 
\be
C_1=\frac{ir(-aQ^2\sin\theta+\rho^2\omega)\cos\theta}{\rho^2KQ\sin\theta}Z_1-\frac{i}{rQ^2\sin\theta}\{Y_2+(\frac{a\omega\cos\theta}{Q}-\frac{(r^2+a^2)\cot \theta}{\rho^2})Z_2\},
\ee
in terms of another new function $Y_2(r,\theta)$. We notice later that the new function $Y_2$ also can be expressed in terms of Teukolsky functions. The function $B_1$ in (\ref{htth1}) and (\ref{hthp1}) and $C_2$ in (\ref{htr1}), (\ref{hrp1}) are given by
\be
B_1=\frac{\Delta}{iK}\{\frac{2ia\cos\theta}{\rho^2}C_1-\frac{\partial}{\partial r}(J+H)\},
\ee
\be
C_2=\frac{1}{Q}\{\frac{-2iar\sin\theta}{\rho^2}B_2+\frac{1}{\sin\theta}\frac{\partial}{\partial \theta}(F-G)\sin\theta\}.
\ee
The relevant combinations of components of ${\bold F}$ that appear in equations (\ref{htt1})-(\ref{htp1}), (\ref{hrp1}) and (\ref{hthp1}) are given by
\be
F_2^1+F_1^2=\frac{\sqrt{2}}{iK}\{\frac{-ia}{rKQ\cos\theta}[r^2Q\sin\theta Z_1-aK\cos^2\theta Z_2]+[k,\nu]+[k,\nu]^*\},
\ee
\be
F_3^4+F_4^3=\frac{\sqrt{2}}{iaQ\rho^2 \sin \theta}\{\frac{ia}{rKQ\cos\theta}[r^2Q\sin\theta Z_1-aK\cos^2\theta Z_2]+[\lambda,\sigma]+[\lambda,\sigma]^*\},
\ee
\be
F_2^1-F_1^2=\frac{\sqrt{2}}{Q}\{-i\frac{K^2\rho^4-2a^2\Delta^2\cos^2\theta}{2\Delta K^2\rho^2\cos^2\theta}Z_1-\frac{ia\Delta\cos\theta}{rKQ\sin\theta}[(\frac{r-M}{\Delta}-\frac{r}{\rho^2})Z_2-X_1]-\frac{\Delta}{2\rho^2}\frac{\partial}{\partial r}\frac{\rho^4}{\Delta}B_2+<k,\nu>\},
\ee
\be
F_4^3-F_3^4=\frac{-\sqrt{2}}{iK\rho^2}\{i\frac{Q^2\rho^4+2a^2r^2\sin^2\theta}{2r\rho^2Q^2\sin^2\theta}Z_2+\frac{iar\sin\theta}{KQ\cos\theta}[\frac{(r^2+a^2)\cot \theta}{\rho^2}Z_1+Y_1]-\frac{1}{2\rho^2}(\frac{\partial}{\partial \theta}-\cot\theta)\rho^4C_1+<\lambda,\sigma>\},
\ee
where $[\kappa,\nu]$ and $<\kappa,\nu>$ are given by 
\be
[\kappa, \nu]=\frac{\rho^4\bar \rho}{2\Delta}\{\frac{{\cal L}_1(\rho^2\nu^{(1)*})}{\rho^2\bar \rho^*}-\frac{ia\sin\theta}{\bar \rho^2}\nu^{(1)}\}+\frac{\Delta}{8}\{\bar \rho^2{\cal L}^\dag _1(\frac{\kappa^{(1)*}}{\bar \rho^2})-\frac{ia\sin\theta}{\bar \rho^*}\kappa^{(1)}\},
\ee
and
\be
<\kappa,\nu>=\frac{1}{2\rho^2}\{\frac{\Delta}{2}(\bar \rho^*\kappa^{(1)*}-\bar \rho\kappa^{(1)})+\frac{2\rho^4}{\Delta}(\bar \rho^*\nu^{(1)}-\bar \rho\nu^{(1)*})\}.
\ee
Moreover, $[\lambda,\sigma]$ and $<\lambda,\sigma>$ stand for 
\be
[\lambda,\sigma]=\frac{(\bar \rho^*)^2}{2\sqrt 2}\{{\cal D}_0(\bar \rho^*\lambda^{(1)})+\frac{\Delta}{2\rho^2}\sigma^{(1)*}\}-\frac{\rho^2}{2\sqrt 2}\{\Delta\frac{\bar \rho^*}{2\bar \rho}{\cal D}^\dag _1 \frac{\bar \rho\sigma^{(1)}}{(\bar \rho^*)^2}+\frac{\bar \rho}{\bar \rho^*}\lambda^{(1)*}\},
\ee
\be
<\lambda,\sigma>=\frac{1}{\sqrt{2}\rho^2}\{\bar \rho^2(\rho ^2\lambda^{(1)*}-\frac{1}{2}\Delta\sigma^{(1)})-(\bar \rho^*)^2(\rho ^2\lambda^{(1)}-\frac{1}{2}\Delta \sigma^{(1)*}) \},
\ee
where the spin coefficients $\kappa^{(1)},\nu^{(1)},\lambda^{(1)}$ and $\sigma^{(1)}$ are given by (\ref{sc1})-(\ref{sc4}), respectively.
We notice that all components of the gravitational perturbation, except (\ref{hrth1}) are combinations of six functions $X_1,X_2,Y_1,Y_2,Z_1,Z_2$. We use some other known relations to determine and fix these six functions. The summation of $Y_1$ and $Y_2$ as well as $X_1$ and $X_2$ are given respectively by 
\be
Y_1+Y_2=[S]^-\cos \theta(F_1-\frac{4rK}{\rho^2}F_2)+\frac{i}{a}[S]^+(F_3-\frac{4rK}{\rho^2}F_4),\, 
\ee
and
\be
X_1+X_2=\frac{r[P]^+}{a\Delta}(G_1-\frac{4a^2Q\cos\theta}{\rho^2}G_2)+\frac{-i[P]^-}{\Delta}(G_3-\frac{4a^2Q\cos\theta}{\rho^2}G_4)\label{X1pX2},\, 
\ee
where the $r$-dependent functions $F_i,\, i=1,\cdots ,4$ are
\begin{eqnarray}
F_1&=&K[DDP]^++2r\omega[DP]^+\label{F1},\\
F_2&=&r[DDP]^+-[DP]^+,\\
F_3&=&3Kr[DDP]^--2\alpha^2\omega[DP]^--6r\omega[P]^-\label{F3},\\
F_4&=&r^2[DDP]^--r[DP]^-\label{F4}.
\end{eqnarray}
The different functions that appear in right hand side of equations (\ref{F1})-(\ref{F4}) are combinations of the radial Teukolsky functions $R_{\pm 2}$ and their derivatives as
\begin{eqnarray}
[P]^\pm &=&P_+\pm P_-\label{Ppm},\\
{[DP]}^\pm &=&{\cal D}_0^\dagger P_+\pm {\cal D}_0 P_-,\\
{[DDP]}^\pm &=&{\cal D}_0^\dagger {\cal D}_0^\dagger P_+\pm {\cal D}_0 {\cal D}_0 P_-,
\end{eqnarray}
where $P_+=\Delta^2 R_{+2},
P_-=R_{-2}$.
Moreover, the $\theta$-dependent functions $G_i,\, i=1,\cdots,4$ in (\ref{X1pX2}) are
\begin{eqnarray}
G_1&=&Q[LLS]^-+2a\omega[LS]^-\cos\theta,\\
G_2&=&[LLS]^-+[LS]^-\sin \theta,\\
G_3&=&3Q[LLS]^+\cos \theta+\frac{2\alpha^2\omega}{a}[LS]^++6a\omega[S]^+\sin \theta\cos\theta,\label{G3}\\
G_4&=&[LLS]^+\cos ^2\theta+[LS]^+\sin \theta\cos \theta,
\end{eqnarray}
where $[S]^{\pm},[LS]^{\pm}$ and $[LLS]^{\pm}$ are combinations of angular Teukolsky functions and their derivatives as
\begin{eqnarray}
{[S]}^\pm &=&S_+\pm S_-,\label{Spm}\\
{[LS]}^\pm &=&{\cal L}_2 S_+\pm {\cal L}_2^\dagger S_-,\\
{[LLS]}^\pm &=&{\cal L}_1{\cal L}_2S_+\pm {\cal L}_1^\dagger {\cal L}_2^\dagger S_-.\label{LLSpm}
\end{eqnarray}
We denote the angular Teukolsky functions $S_{\pm 2}$ by $S_{\pm}$ in (\ref{Spm})-(\ref{LLSpm}) to simplify the notation. The constant $\alpha$ in equations (\ref{F3}) and (\ref{G3}) is $
\alpha^2=a^2+\frac{am}{\omega}
$.
A straightforward calculation shows that
\bea
[DP]^\pm&=&\frac{-iK}{\Delta}[P]^\mp+\frac{d}{dr}[P]^\pm\label{DP},\\
{[DDP]}^\pm&=&\frac{-iK}{\Delta}[DP]^\mp+\frac{d}{dr}{[DP]}^\pm\nn\\
&=&\frac{-iK}{\Delta}(\frac{-iK}{\Delta}[P]^\pm+\frac{d}{dr}[P]^\mp)+\frac{d}{dr}(\frac{-iK}{\Delta}[P]^\mp+\frac{d}{dr}[P]^\pm),\label{DDP}\\
{[LS]}^\pm&=&Q[S]^\mp+2[S]^\pm\cot \theta+\frac{d}{d\theta}[S]^\pm,\label{LS}\\
{[LLS]}^\pm&=&Q[LS]^\mp+[LS]^\pm\cot \theta+\frac{d}{d\theta}{[LS]}^\pm\nn\\
&=&Q(Q[S]^\pm+2[S]^\mp\cot \theta+\frac{d}{d\theta}[S]^\mp)+(Q[S]^\mp+2[S]^\pm\cot \theta+\frac{d}{d\theta}[S]^\pm)\cot \theta\nn\\&+&\frac{d}{d\theta}(Q[S]^\mp+2[S]^\pm\cot \theta+\frac{d}{d\theta}[S]^\pm).\label{LLS}
\eea
Moreover, we have the following equations
\be
X_1-X_2=-\frac{D+\Gamma _1}{96M\Delta\sqrt{2}}(-i[P]^-+\frac{4\bar\lambda\omega}{D+\Gamma_1}r[P]^+){\cal I},\label{X1mX2}
\ee
and
\be
Y_1-Y_2=\frac{D+\Gamma_1}{96M\sqrt{2}}\{[S]^-\cos\theta+\frac{4\omega(\bar\lambda \alpha^2-6a^2)}{a(D+\Gamma_1)}[S]^+\}{\cal R},\label{Y1mY2}
\ee
where
\bea
{\cal I}&=&\frac{1}{-\bar \lambda a \omega \beta_1}\{(2(Q\cot\theta-a\omega\cos\theta)(D-\Gamma_1)+8a\omega\bar \lambda Q^2\cos\theta)[S]^+\nn\\&+&(-2Q^2(D-\Gamma_1)-8a\omega\bar \lambda(\frac{Q}{\sin\theta}-a\omega\cos^2\theta))[S]^-\} ,\label{III}
\eea
and
\bea
{\cal R}&=&\frac{4}{B_1(D+\Gamma_1)}\{(2K^2(D+\Gamma_1)+8\bar \lambda\omega r(K(r-M)-r\omega\Delta)-8\bar \lambda \omega K \Delta)[P]^+\nn\\&+&(-8K^2\bar \lambda \omega r+2(D+\Gamma_1)(K(r-M)-\omega r \Delta))(-i[P]^-)\} .\label{RRR} 
\eea
In equations (\ref{X1mX2})-(\ref{RRR}), $D=\sqrt{\bar \lambda ^2-4\alpha^2\omega^2}$ is the Starobinsky constant and $\Gamma_1=\bar \lambda(\bar \lambda +2)-12\alpha^2\omega^2$. 
So, as we notice, we can find the functions $
X_1,X_2,Y_1,Y_2$ in terms of functions $[P]^\pm$ and $[S]^\pm$ and their first and second derivatives. 
Moreover, we are able to find the first and second derivatives $\frac{dP_\pm}{dr},\frac{d^2P_\pm}{dr^2},\frac{dS_\pm}{d\theta}$ and $\frac{d^2S_\pm}{d\theta^2}$ in terms of Teukolsky functions. The first derivatives of radial Teukolsky functions are given by
\bea
\frac{dP_+}{dr}&=&\frac{-i}{A_3}\{\Delta O P_+-((A_1-\frac{A_3K}{\Delta})+iA_2)P_-\},\\
\frac{dP_-}{dr}&=&\frac{i}{A_3}\{\Delta O^* P_--((A_1-\frac{A_3K}{\Delta})-iA_2)P_+\},\\
\eea
where
\bea
O&=&D+iO_2=D-i\Gamma_2=D-12iM\omega,\\
A_1&=&\Delta \Gamma_1-4\bar \lambda K^2+24\omega K (a^2-Mr),\\
A_2&=&-\Delta \Gamma_2-4\bar \lambda (K(r-M)+r\Delta \omega)+24\omega r K^2,\\
A_3&=&-8K^3+8K(a^2-M^2)-8\omega \Delta (a^2-M r)+4\bar \lambda K \Delta.
\eea
Moreover, the derivatives of angular Teukolsky functions are given by
\bea
\frac{dS_+}{d\theta}&=&-(2\cot\theta+Q+\frac{\alpha_1+\alpha_2}{\beta_1})S_++\frac{D}{\beta_1}S_-,\\
\frac{dS_-}{d\theta}&=&-(2\cot\theta-Q+\frac{-\alpha_1+\alpha_2}{\beta_1})S_--\frac{D}{\beta_1}S_+,
\eea
where the angular functions $\alpha_1,\alpha_2$ and $\beta_1$ are
\bea
\alpha_1&=&\bar \lambda (\bar \lambda+2)-12\alpha^2\omega^2+24\frac{a\omega Q}{\sin\theta}-
4\bar\lambda Q^2,\\
\alpha_2&=&-24a\omega Q^2 \cos\theta+4\bar \lambda(Q\cot\theta+a\omega\cos\theta),\\
\beta _1&=& 8Q^3-8Q\cot ^2 \theta-4(\bar \lambda+2)Q+8a\frac{\omega}{sin\theta}.
\eea
We can also find the second derivatives of Teukolsky functions in terms of Teukolsky functions. As an example, we find that the second derivative of radial Teukolsky function $P_+$ is
\bea
&&\frac{d^2P_+}{dr^2}=-iO\Delta\{\frac{1}{A_3'}-\frac{\Delta'}{A_3}-\frac{\Delta A_2}{A_3^2}\}P_+\nonumber\\&+&\{\frac{\Delta^2(D^2+O_2^2)}{A_3^2}+\frac{i}{A_3'}((A_1-\frac{A_3K}{\Delta})+iA_2)-\frac{i}{A_3}((A_1-\frac{A_3K}{\Delta})'+iA_2')-\frac{1}{A_3^2}((A_1-\frac{A_3K}{\Delta})+iA_2)^2\}P_-.\nonumber\\
&&
\eea
Using the first and second derivatives of the Teukolsky functions, we can express the functions $\frac{d[P]^\pm}{dr},\frac{d^2[P]^\pm}{dr^2},\frac{d[S]^\pm}{d\theta}$ and $\frac{d^2[S]^\pm}{d\theta^2}$ that appear in $
X_1,X_2,Y_1,Y_2$ entirely in terms of Teukolsky functions $P_\pm$ and $S_\pm$. 
Moreover, the functions 
$Z_1$ and $Z_2$ are given by
\be
Z_i= \frac{Z_{i1}+Z_{i2}}{Z_{i0}}, \,\,\,
\ee
where
\be
Z_{11}=-\left( D+{\Gamma_1} \right) \frac{aK\cos \theta }{4Q}
\left( {\cal R}\, \left( {[S]^-}\,\cos \theta +4\,{
\frac {{[S]^+}\,\omega\, \left( \bar \lambda\,{\alpha}^{2}-6\,{a}^{2}
\right) }{ \left( D+{\Gamma_1} \right) a}} \right) -3\,{\frac {a
\omega\, {\cal R}\,{\cal I}\cos \theta}{Q}} \right) ,
\ee
\bea
Z_{12}&=&\nn\\
&K&\cos \theta \left[ \left( {[LS]^-}- \left( \cot
\theta +3\,{\frac {a\omega\,\cos \theta
}{Q}} \right) {[S]^-}+Q{[S]^+} \right) \left( \left( 
{r}^{2}-{a}^{2} \cos^2 \theta
\right) {[DDP]^+}-2\,r{[DP]^+} \right) \right. \nn \\
&-&\left. 2ia \left( {[LS^+]
}- \left( {\frac {1}{\sin \theta \cos \theta
}}+3\,{\frac {a\omega\,\cos\theta }{Q}}
\right) {[S]^+}+Q{[S]^-} \right)\left( r{[DDP]^-}-{[DP]^-} \right) \cos \theta \right. \nn\\&+&\left.2\,{a}^{2}{[DDP]^+}\,{[S]^-}\,\sin \theta \cos \theta 
\right],
\eea
\be
Z_{10}={\rho}^{2} \left( {Q}^{2}- \frac{1}{\sin^2\theta} +\frac{a\omega}{Q\sin\theta}\, \left( 3\, \sin^2 \theta 
-2+3\,{\frac {a\omega\, \cos^2 \theta \sin \theta }{Q}} \right) \right),
\ee
\be
Z_{20}={\rho}^{2} \left( {\frac {-{K}^{2}-{M}^{2}+{a}^{2}}{\Delta}}-\frac{\omega\,
\Delta}{K}\left( 1+2\,{\frac {r \left( r-M \right) }{\Delta}}-3\,{
\frac {{r}^{2}\omega}{K}} \right) \right), 
\ee
\be
Z_{21}=\left( D+{\Gamma_1} \right) \frac{arQ}{4K}\Delta
\, {\cal I} \sin\theta\left( \, \left( -i[P]^-+4\,{\frac {\lambda\,\omega\,r[P]^+}{D+
{\Gamma_1}}} \right) \frac{1}{\Delta}-3\,{\frac {r\omega\,{\cal R}}{K}} \right),
\ee
\bea
Z_{22}&=&\frac{rQ}{a}\left[ \left( {[DP]^+}- \left( {\frac {r-M}{\Delta}}+3\,{\frac 
{r\omega}{K}} \right) [P]^+-{\frac {iK[P]^-}{\Delta}} \right) \right. \nn \\ 
&\times&\left. \left( \left( {
r}^{2}-{a}^{2} \cos^2\theta 
\right) {[LLS]^-}-2\,{a}^{2}{[LS]^-}\,\sin \theta
\cos\theta \right) \right.\nn\\
&-&\left.2ar \left( i{[DP]^-
}-i \left( {\frac {Mr-{a}^{2}}{r\Delta}}+3{\frac {r\omega}{K}}
\right) [P]^-+{\frac {K}{\Delta}[P]^+} \right) \right. \nn\\ 
&\times&\left. \left( {[LLS]^+}\cos
\theta +{[LS]^+}\sin \theta 
\right) +2r{[LLS]^-}[P]^+ \right].
\eea
We note that using equations (\ref{DP})-(\ref{LLS}), (\ref{Ppm}) and (\ref{Spm}), we can express the functions $Z_i,\,i=1,2$ completely in terms of $P_{\pm}$ and $S_{\pm}$. The expressions are very long and so we do not present them here. 
Moreover, we find that all the spin coefficients (\ref{sc1})-(\ref{sc4}) can be written in terms of Teukolsky functions $R_{\pm}$ and $S_{\pm}$ 
\bea
\kappa^{(1)}&=&\frac{-\sqrt 2 (a\cos\theta+ir) R_+S_+}{6M\sin \theta \beta_1}\{(a\sin\theta \cos \theta+ir\sin\theta) ({\alpha_1}+\alpha_2)+3 a\cos ^2\theta 
{\beta_1}-3a{\beta_1}
\}\nn\\
&+&\frac{\sqrt 2 D (a\cos\theta+ir)^2R_+S_-}{6M\beta_1},
\eea

\bea
\sigma^{(1)}&=& -\frac{S_{+2}R_+(ir+a\cos\theta)}{6A_3M(ir-a\cos\theta)}\{{\Delta}^{2}aO\cos \theta +i{\Delta}^{2}{O}r+A_3(-4iMa
\cos \theta +4iar\cos \theta 
\nn\\
&+&4Mr+iKr+Ka\cos \theta-4{r}^{2}+3\Delta)
\}\nn\\
&-&\frac{R_-S_+}{6A_3\Delta M(a\cos\theta-ir)}(\Delta A_1-KB_1+iA_2\Delta)(a\cos\theta+ir)^2,
\eea
\bea
\lambda^{(1)}&=&\frac{S_-R_-}{3M A_3 \Delta (r-ia\cos\theta)^2}\{\Delta ^2 a O^* \cos\theta+KaA_3\cos\theta+i\Delta^2O^*r+irKA_3-3\Delta A_3\}\nonumber\\
&+&\frac{S_-R_+}{3A_3M(ir+a\cos\theta)}\{\Delta (A_1-iA_2)-KA_3\},
\eea
\bea
\nu^{(1)}&=&\frac{\sqrt 2 R_-S_-}{6M\sin \theta \beta_1 \rho^2 (ir+a\cos\theta)}\{(a\sin\theta \cos \theta+ir\sin\theta) ({\alpha_1}-\alpha_2)-3 a\cos ^2\theta 
{\beta_1}+3a{\beta_1}
\}\nn\\
&-&\frac{\sqrt 2 D R_-S_+}{M\beta_1\rho^2}.
\eea
Furnished by all necessary functions that appear in gravitational perturbation, we find that 
all components of the gravitational field (\ref{htt1})-(\ref{hthp1}) can be written as
\be
h^{\mu \nu}=\sum _{i,j=+,-} f^{\mu\nu}_{ij}(r,\theta) P_i (r) S_j (\theta)\label{hps},
\ee
where 
\be
f^{\mu\nu}_{ij}=f^{\mu\nu(-1)}_{ij}\Delta^{-1}+f^{\mu\nu(0)}_{ij}\Delta^{0}+\cdots\label{compo}.
\ee
As an example, we find that the functions $f^{rr(-1)}_{ij}$ and $f^{rr(0)}_{ij}$ are
\bea
f^{rr(-1)}_{++}&=&\frac{2{\sqrt{2}}ar 
\sin^2 \theta Q{{
A_3}}^{2}K}{{\cal F}}
\{
12M\sin \theta \cos^2\theta {a}^{2}{\beta_1}\omega r+3K
\sin \theta \cos \theta a{\beta_1}
\omega{r}^{2}-4M\sin \theta \cos \theta
{Q}^{2}a{\beta_1}r\nn\\
&+&4 D M\sin 
\theta \cos \theta Qar+4M\sin \theta
\cos\theta Qa{\alpha_1}r+4M\sin 
\theta \cos\theta Qa{\alpha_2}r+K\sin
\theta Q{\alpha_2}{r}^{2}\nn\\
&-&2K\cos \theta
Q{a}^{2}{\beta_1}+K\cos \theta Q{\beta_1}
{r}^{2}+4MQa{\beta_1}r+K \cos ^3\theta 
Q{a}^{2}{\beta_1}-K\sin \theta {Q}^{2}{
\beta_1}{r}^{2}- D K\sin \theta Q{r}
^{2}\nn\\
&+&K\sin \theta Q{\alpha_1}{r}^{2}-K\sin 
\theta \cos^2 \theta Q{a}^{2
}{\alpha_2}+4\sin \theta \cos \theta
{Q}^{2}a{\beta_1}{r}^{2}-4 D \sin
\theta \cos\theta Qa{r}^{2}\nn\\
&-&4\sin
\theta \cos \theta Qa{\alpha_1}{r}^
{2}
-4\sin \theta \cos \theta Qa{\alpha_2}{r}^{2}
+2iKQa{\beta_1}r-3K\sin \theta 
\cos ^3 \theta {a}^{3}{\beta_1}
\omega \nn\\
&+&K\sin \theta \cos ^2\theta 
{Q}^{2}{a}^{2}{\beta_1}-12\sin \theta 
\cos ^2\theta {a}^{2}{\beta_1}
\omega{r}^{2}+ D K\sin \theta 
\cos ^2 \theta Q{a}^{2}-K\sin \theta
\cos ^2\theta Q{a}^{2}{\alpha_1}\nn\\
&-&4Qa{\beta_1}{r}^{2}+6iK\sin \theta 
\cos ^2\theta {a}^{2}{\beta_1}
\omega r-2iK\sin \theta \cos\theta {
Q}^{2}a{\beta_1}r+2i DK\sin \theta
\cos \theta Qar\nn\\
&+&2iK\sin\theta
\cos \theta Qa{\alpha_1}r+2iK\sin
\theta \cos \theta Qa{\alpha_2}r
\},\label{firstf}
\eea
where ${\cal F}$ is given by
\bea
{\cal F}&=&{\rho}^{4}K{\beta_1}{{A_3}}^{2} \left( 3{a}^{2} \cos
^4\theta {\omega}^{2}+3\sin\theta
\cos ^2\theta Qa\omega+{Q}^
{4} \cos^2 \theta -3 \cos
^2\theta {a}^{2}{\omega}^{2}-\sin 
\theta Qa\omega-{Q}^{4}+{Q}^{2} \right), \nn\\
&&
\eea
and
\bea
f^{rr(0)}_{++}&=&\frac{
-2{\sqrt{2}}ar\sin ^2 \theta Q{A_3}}{{\cal F}} \{ 4{A_3}Qa{\beta_1}\omega{r}^{2}+2{
A_2}KQa{\beta_1}r+2{A_3}K\sin \theta \cos
\theta Qa{\alpha_1}+2{A_3}K\sin \theta
\cos \theta Qa{\alpha_2}\nn\\
&-&i{A_2}K
\cos ^{3}\theta Q{a}^{2}{\beta_1}
-i{
A_2}K\sin \theta Q{\alpha_1}{r}^{2}-i{A_2}
K\sin \theta Q{\alpha_2}{r}^{2}+2i{A_2}K
\cos \theta Q{a}^{2}{\beta_1}
-i{ A_2}K\cos
\theta Q{\beta_1}{r}^{2}\nn\\
&+&4i{A_1}KQa{\beta_1}r+12{A_3}\sin \theta \cos^{2}
\theta {a}^{2}{\beta_1}{\omega}^{2}{r}^{2}
+i{A_2}K\sin \theta {Q}^{2}{\beta_1}{r}^{
2}
+6{A_3}K\sin \theta \cos ^{2}\theta
{a}^{2}{\beta_1}\omega\nn\\
&+&i{A_2} D
K\sin\theta Q{r}^{2}-2{A_3}K\sin
\theta \cos \theta {Q}^{2}a{\beta_1}+
2{A_3} D K\sin \theta \cos
\theta Qa+i{A_2}K\sin\theta 
\cos^{2} \theta Q{a}^{2}{\alpha_2}\nn\\
&+&6
{A_2}K\sin \theta \cos ^{2}\theta
{a}^{2}{\beta_1}\omega r-4{A_3}\sin
\theta \cos \theta {Q}^{2}a{\beta_1}
\omega{r}^{2}-2{A_2}K\sin \theta \cos
\theta {Q}^{2}a{\beta_1}r\nn\\
&+&4{A_3} D
\sin \theta \cos \theta Qa
\omega{r}^{2}
+4{A_3}\sin \theta \cos
\theta Qa{\alpha_1}\omega{r}^{2}+4{A_3}\sin
\theta \cos \theta Qa{\alpha_2}
\omega{r}^{2}+2{A_2} D K\sin \theta
\cos\theta Qar\nn\\
&+&2{A_2}K\sin 
\theta \cos \theta Qa{\alpha_1}r+2{A_2
}K\sin \theta \cos \theta Qa{\alpha_2}r+3i{A_2}K\sin \theta \cos^{3}
\theta{a}^{3}{\beta_1}\omega\nn\\
&-&i{A_2}K\sin \theta\cos ^{2}\theta 
{Q}^{2}{a}^{2}{\beta_1}-iA_2 D K
\sin \theta \cos ^{2}\theta Q{a}^{2}+12i{A_1}K\sin \theta 
\cos ^{2}\theta {a}^{2}{\beta_1}
\omega r\nn\\
&-&4i{A_1}K\sin \theta \cos \theta
{Q}^{2}a{\beta_1}r-3i{A_2}K\sin \theta
\cos \theta a{\beta_1}\omega{r}^{2}+4i
{A_1} D K\sin \theta \cos 
\theta Qar\nn\\
&+&4i{A_1}K\sin \theta \cos
\theta Qa{\alpha_1}r+4i{A_1}K\sin \
\theta \cos \theta Qa{\alpha_2}r+i{A_2}
K\sin \theta \cos ^{2}\theta 
Q{a}^{2}{\alpha_1}+2{A_3}KQa{\beta_1} \},
\eea
\bea
f^{rr(-1)}_{-+}&=&
\frac{2{\sqrt 2}ar \sin^{2} \theta Q{{A_3}}^{2}K}{{\cal F}}\{-4Qa{\beta_1}{r}^{2}+4MQa{\beta_1}r+K
\cos^{3} \theta Q{a}^{2}{\beta_1}-K
\sin \theta {Q}^{2}{\beta_1}{r}^{2}- D
K\sin\theta Q{r}^{2}\nn\\
&+&K\sin \theta
Q{\alpha_1}{r}^{2}+K\sin \theta Q{\alpha_2}{r}^{2}-2K\cos \theta Q{a}^{2}{\beta_1}+K
\cos \theta Q{\beta_1}{r}^{2}-3K\sin 
\theta \cos^3 \theta {a}^{3}
{\beta_1}\omega\nn\\
&+&K\sin \theta \cos ^2
\theta {Q}^{2}{a}^{2}{\beta_1}-12\sin 
\theta \cos ^{2} \theta {a}^{2}
{\beta_1}\omega{r}^{2}+ D K\sin \theta
\cos ^{2}\theta Q{a}^{2}-K
\sin\theta \cos^{2}\theta 
Q{a}^{2}{\alpha_1}\nn\\
&-&K\sin\theta 
\cos ^{2}\theta Q{a}^{2}{\alpha_2}+4\sin
\theta \cos \theta {Q}^{2}a{\beta_1}
{r}^{2}-4 D \sin \theta \cos 
\theta Qa{r}^{2}
-4\sin \theta \cos 
\theta Qa{\alpha_1}{r}^{2}\nn\\
&-&4\sin \theta 
\cos \theta Qa{\alpha_2}{r}^{2}-2iKQa{\beta_1}
r-6iK\sin \theta \cos ^{2} \theta
{a}^{2}{\beta_1}\omega r+2iK\sin 
\theta \cos \theta {Q}^{2}a{\beta_1}r\nn\\
&-&2i
D K\sin \theta \cos \theta
Qar-2iK\sin \theta \cos \theta
Qa{\alpha_1}r-2iK\sin \theta \cos
\theta Qa{\alpha_2}r+12M\sin \theta
\cos^{2} \theta {a}^{2}{\beta_1}\omega r\nn\\
&+&3K\sin \theta \cos \theta
a{\beta_1}\omega{r}^{2}-4M\sin \theta 
\cos \theta {Q}^{2}a{\beta_1}r+4 D
M\sin \theta \cos \theta Qar+4
M\sin \theta \cos \theta Qa{\alpha_1
}r\nn\\
&+&4M\sin \theta \cos\theta Qa{\alpha_2}r \},
\eea
\bea
f^{rr(0)}_{-+}&=&\frac{
2{\sqrt2}ar \sin ^{2} \theta Q{A_3}}{{\cal F}} \{-4{A_3} D \sin \theta
\cos\theta Qa\omega{r}^{2}-4{A_3}
\sin \theta \cos \theta Qa{\alpha_1}
\omega{r}^{2}-4{ A_3}\sin \theta \cos 
\theta Qa{\alpha_2}\omega{r}^{2}\nn\\
&-&2{ A_2} D
K\sin \theta \cos \theta Qar-2
{A_2}K\sin \theta \cos \theta Qa{
\alpha_1}r-2{A_2}K\sin \theta \cos 
\theta Qa{\alpha_2}r+3i{ A_2}K\sin \theta
\cos ^{3}\theta {a}^{3}{\beta_1}\omega\nn\\
&-&i{A_2}K\sin \theta \cos^{2}
\theta {Q}^{2}{a}^{2}{\beta_1}-i{A_2
} D K\sin \theta \cos ^2
\theta Q{a}^{2}+i{A_2}K\sin \theta
\cos ^2\theta Q{a}^{2}{ 
\alpha_1}\nn\\
&+&i{A_2}K\sin \theta \cos ^{2}
\theta Q{a}^{2}{\alpha_2}-6{A_2}K\sin
\theta \cos ^{2}\theta 
{a}^{2}{\beta_1}\omega r+4{A_3}\sin \theta
\cos \theta {Q}^{2}a{\beta_1}\omega{r}^{
2}\nn\\
&+&2{A_2}K\sin \theta \cos \theta
{Q}^{2}a{\beta_1}r-4{A_3}Qa{\beta_1}\omega{r
}^{2}-2{A_2}KQa{\beta_1}r-2{A_3}KQa{\beta_1}-12{
A_3}\sin \theta \cos^2 \theta
{a}^{2}{\beta_1}{\omega}^{2}{r}^{2}\nn\\
&+&i{A_2
}K\sin \theta {Q}^{2}{\beta_1}{r}^{2}-6{A_3}
K\sin \theta \cos^2 \theta 
{a}^{2}{\beta_1}\omega+i{A_2} D K
\sin \theta Q{r}^{2}+2{A_3}K\sin \theta
\cos \theta {Q}^{2}a{\beta_1}\nn\\
&-&2{A_3}
DK\sin \theta \cos \theta
Qa-2{A_3}K\sin \theta \cos 
\theta Qa{\alpha_1}-2{A_3}K\sin \theta
\cos \theta Qa{\alpha_2}-i{A_2}K
\cos^3 \theta Q{a}^{2}{\beta_1}\nn\\
&-&i{
A_2}K\sin \theta Q{\alpha_1}{r}^{2}-i{A_2}
K\sin \theta Q{\alpha_2}{r}^{2}+2i{A_2}K
\cos \theta Q{a}^{2}{\beta_1}-i{A_2}K\cos
\theta Q{\beta_1}{r}^{2}+4i{A_1}KQa{\beta_1}r\nn\\
&-&4i{A_1}K\sin \theta \cos \theta
{Q}^{2}a{\beta_1}r-3i{A_2}K\sin \theta
\cos \theta a{\beta_1}\omega{r}^{2}+4i
{A_1} D K\sin \theta \cos 
\theta Qar\nn\\
&+&4i{A_1}K\sin \theta \cos
\theta Qa{\alpha_1}r
+4i{A_1}K\sin 
\theta \cos \theta Qa{\alpha_2}r+12i{
A_1}K\sin \theta \cos ^{2}\theta 
{a}^{2}{\beta_1}\omega r \},
\eea
\bea
f_{--}^{rr(-1)}&=&-\frac{{\sqrt 2}ar \sin ^2\theta Q}{{\cal F}}
\{ 4ia{{\it A_3}}^{2}{K}^{2}\cos \theta \sin
\theta {Q}^{2}{\beta_1}r+2{{A_3}}^{2}{K}^{2}
\cos \theta Q{\beta_1}{r}^{2}+2\sin \theta
{{ A_3}}^{2}{K}^{2}{Q}^{2}{\beta_1}{r}^{2}\nn\\
&+&8a{{A_3
}}^{2}KQ{\beta_1}{r}^{2}-2\sin \theta {{A_3}}^
{2}{K}^{2}Q{\alpha_1}{r}^{2}+2\sin \theta {{ 
A_3}}^{2}{K}^{2}Q{\alpha_2}{r}^{2}+2\sin \theta {{A_3}}^{2} D {K}^{2}Q{r}^{2}\nn\\
&-&4\cos \theta
{a}^{2}{\beta_1}Q{{A_3}}^{2}{K}^{2}+2{{A_3}}^{2}{
K}^{2} \cos ^3\theta Q{a}^{2}{\beta_1}-8a{{A_3}}^{2}KM\cos \theta \sin 
\theta {Q}^{2}{\beta_1}r\nn\\
&-&8a{{A_3}}^{2}KM\cos 
\theta \sin \theta Q{\alpha_2}r-24{a}^{2}
{{A_3}}^{2}KM \cos^2 \theta \sin
\theta {\beta_1}\omega r+8a{{A_3}}^{2}
D KM\cos \theta \sin \theta
Qr\nn\\
&+&8a{{A_3}}^{2}KM\cos\theta \sin
\theta Q{\alpha_1}r-4ia{{A_3}}^{2}{K}^{2}\cos
\theta \sin \theta Q{\alpha_1}r-4
ia{{A_3}}^{2} D {K}^{2}\cos \theta 
\sin \theta Qr\nn\\
&+&4ia{{ A_3}}^{2}{K}^{2}Q{\beta_1}
r+6\sin \theta {{A_3}}^{2}{K}^{2}\cos 
\theta a{\beta_1}\omega{r}^{2}-8a{{A_3}}^{2}
D K\cos \theta \sin \theta
Q{r}^{2}-8a{{A_3}}^{2}K\cos \theta \sin
\theta Q{\alpha_1}{r}^{2}\nn\\
&+&8a{{A_3}}^{2}K\cos
\theta \sin \theta Q{\alpha_2}{r}^{
2}+8a{{ A_3}}^{2}K\cos \theta \sin\theta
{Q}^{2}{\beta_1}{r}^{2}+24{a}^{2}{{ A_3}}^{2}K
\cos ^2\theta\sin \theta
{\beta_1}\omega{r}^{2}\nn\\
&-&2\sin \theta {{
A_3}}^{2}{K}^{2} \cos ^2 \theta Q{a
}^{2}{\alpha_2}-6\sin \theta {{A_3}}^{2}{K}^{2}
\cos ^3 \theta {a}^{3}{\beta_1}
\omega-8a{{A_3}}^{2}KMQ{\beta_1}r\nn\\
&-&2\sin \theta
{{A_3}}^{2} D {K}^{2} \cos ^2
\theta Q{a}^{2}-2\sin \theta {{
A_3}}^{2}{K}^{2} \cos ^2 \theta{Q}
^{2}{a}^{2}{\beta_1}
+2\sin \theta {{A_3}}^{2}{K}
^{2}\cos ^2 \theta Q{a}^{2}{\alpha_1}\nn\\
&+&4ia{{A_3}}^{2}{K}^{2}\cos \theta \sin
\theta Q{\alpha_2}r+12i{a}^{2}{{A_3}}^{2}{K}^
{2} \cos ^2\theta \sin \theta
{\beta_1}\omega r 
\},
\eea
\bea
f_{--}^{rr(0)}&=&\frac{
{\sqrt 2}ar \sin ^2\theta Q}{{\cal F}}
\{4i\cos\theta {a}^{2}{\beta_1}Q{ A_2}
{A_3}K+2i\sin \theta {A_2}{A_3}KQ{\alpha_1}{r}^{2}+8a{{A_3}}^{2}\cos \theta \sin
\theta Q{\alpha_2}\omega{r}^{2}\nn\\
&-&8a{{A_3}}^{
2}\cos \theta \sin \theta Q{\alpha_1}
\omega{r}^{2}-8a{{A_3}}^{2} D \cos 
\theta \sin \theta Q\omega{r}^{2}-4a{A_2
}{A_3}K\cos \theta \sin \theta Q{
\alpha_1}r\nn\\
&+&4a{A_2}{A_3}K\cos \theta
\sin \theta Q{\alpha_2}r-4a{A_2}{ A_3}
D K\cos \theta \sin \theta
Qr+2i\sin\theta {A_2}{A_3}
D K \cos ^2 \theta Q{a
}^{2}\nn\\
&-&2i\sin \theta {A_2}{A_3}K{Q}^{2}{\beta_1}{r}^{2}
-2i{A_2}{A_3}K\cos\theta Q{
\beta_1}{r}^{2}-2i\sin \theta {A_2}{A_3}
KQ{\alpha_2}{r}^{2}-2i{A_2}{A_3}K \cos
^3\theta Q{a}^{2}{\beta_1}\nn\\
&+&4a{A_2}
{A_3}K\cos \theta \sin \theta {Q}^{
2}{\beta_1}r+12{a}^{2}{A_2}{A_3}K \cos 
^2\theta \sin \theta {\beta_1}
\omega r+2i\sin \theta {A_2}{A_3}K 
\cos ^2\theta {Q}^{2}{a}^{2}{\beta_1}\nn\\
&+&2i
\sin \theta { A_2}{ A_3}K \cos ^2
\theta Q{a}^{2}{\alpha_2}+6i\sin 
\theta { A_2}{ A_3}K \cos ^3 \theta
{a}^{3}{ \beta_1}\omega-8ia{ A_1}{ 
A_3}KQ{ \beta_1}r-2i\sin \theta { A_2}{ A_3
} D KQ{r}^{2}\nn\\
&+&8ia{ A_1}{ A_3} D
K\cos \theta \sin \theta Qr+8
ia{ A1}{ A_3}K\cos \theta \sin \theta
Q{ \alpha_1}r-2i\sin \theta { A_2}{
A_3}K\cos ^2\theta Q{a}^{2}{
\alpha_1}\nn\\
&+&12{a}^{2}{{ A_3}}^{2}K \cos ^2\theta
\sin \theta { \beta_1}\omega+
24{a}^{2}{{ A_3}}^{2} \cos ^2 \theta \sin \theta { \beta_1}{\omega}^{2}{r}^{2}-4a{{
A_3}}^{2}K\cos \theta \sin \theta Q{
\alpha_1}\nn\\
&+&4a{{ A_3}}^{2}K\cos \theta \sin 
\theta Q{ \alpha_2}+4a{{ A_3}}^{2}K\cos \theta
\sin \theta {Q}^{2}{ \beta_1}+4a{ A_2}{
A_3}KQ{ \beta_1}r-4a{{ A_3}}^{2} D K\cos
\theta \sin \theta Q\nn\\
&+&8a{{ A_3}}^{2}
Q{ \beta1}\omega{r}^{2}+8a{{ A_3}}^{2}\cos \theta
\sin \theta {Q}^{2}{ \beta_1}\omega{r}^{2
}-24i{a}^{2}{ A_1}{ A_3}K \cos ^2\theta
\sin \theta { \beta_1}\omega r\nn\\
&-&8ia{
A_1}{ A_3}K\cos \theta \sin \theta
Q{ \alpha_2}r-6i\sin \theta { A_2}{
A_3}K\cos \theta a{ \beta_1}\omega{r}^{2}
-8
ia{ A_1}{ A_3}K\cos \theta\sin \theta
{Q}^{2}{ \beta_1}r\nn\\
&+&4a{{ A_3}}^{2}KQ{ \beta_1}
\},
\eea
\bea
f_{+-}^{rr(-1)}&=&\frac{2{ \sqrt2}ar \sin ^{2} \theta Q{{
A_3}}^{2}K}{{\cal F}}\{ 3K\sin \theta \cos^3
\theta {a}^{3}{ \beta_1}\omega+K\sin
\theta \cos ^2 \theta 
{Q}^{2}{a}^{2}{ \beta_1}-12\sin \theta \cos^2
\theta {a}^{2}{ \beta_1}\omega{r}^{2
}\nn\\
&+& D K\sin \theta \cos ^2
\theta Q{a}^{2}-K\sin \theta 
\cos^2 \theta Q{a}^{2}{ \alpha_1}+K
\sin \theta \cos ^2\theta 
Q{a}^{2}{ \alpha_2}-4\sin \theta \cos
\theta {Q}^{2}a{ \beta_1}{r}^{2}\nn\\
&+&4 D
\sin \theta \cos \theta Qa{r}^{
2}+4\sin \theta \cos \theta Qa{ 
\alpha_1}{r}^{2}-4\sin \theta \cos \theta
Qa{ \alpha_2}{r}^{2}+2iKQa{ \beta_1}r\nn\\
&+&6iK\sin
\theta \cos^2 \theta 
{a}^{2}{ \beta_1}\omega r+2iK\sin \theta \cos
\theta Qa{ \alpha_2}r+2iK\sin \theta
\cos \theta {Q}^{2}a{ \beta_1}r-2i
D K\sin \theta \cos \theta
Qar\nn\\
&-&2iK\sin \theta \cos \theta
Qa{ \alpha_1}r+4MQa{ \beta_1}r-K \cos ^3 
\theta Q{a}^{2}{ \beta_1}-K\sin \theta
{Q}^{2}{ \beta_1}{r}^{2}- D K\sin 
\theta Q{r}^{2}\nn\\
&+&K\sin \theta Q{ \alpha_1}{r}
^{2}-K\sin \theta Q{ \alpha_2}{r}^{2}+2K\cos
\theta Q{a}^{2}{ \beta_1}-K\cos \theta
Q{ \beta_1}{r}^{2}-4Qa{ \beta_1}{r}^{2}\nn\\
&+&12M\sin
\theta \cos ^2 \theta 
{a}^{2}{ \beta_1}\omega r-3K\sin \theta \cos
\theta a{ \beta_1}\omega{r}^{2}+4M\sin 
\theta \cos \theta {Q}^{2}a{ \beta_1}r-4
D M\sin \theta \cos \theta
Qar\nn\\
&-&4M\sin \theta \cos \theta
Qa{ \alpha_1}r+4M\sin \theta \cos 
\theta Qa{ \alpha_2}r \},
\eea
\bea
f_{+-}^{rr(0)}&=&\frac{2\,i {A_3}\,Qar{\sqrt2} \sin 
\theta ^{2}}{{\cal F}}\{ -4i{A_3} D \sin \theta
\cos \theta Qa\omega{r}^{2}-4i{A_3}
\sin \theta \cos \theta Qa{\alpha_1}
\omega{r}^{2}-2i{A_2} D K\sin \theta
\cos \theta Qar\nn\\
&-&2i{A_2}K\sin 
\theta \cos \theta Qa{\alpha_1}r+2i{\it 
A_2}K\sin \theta \cos \theta Qa{\alpha_2}r+4i{A_3}\sin \theta \cos \theta
{Q}^{2}a{\beta_1}\omega{r}^{2}\nn\\
&+&4i{A_3}\sin
\theta \cos \theta Qa{\alpha_2}
\omega{r}^{2}+6i{\it A_2}K\sin \theta \cos
\theta ^{2}{a}^{2}{\beta_1}\omega r+2i
{A_2}K\sin \theta \cos \theta {Q}^{
2}a{\beta_1}r-4{A_1}KQa{\beta_1}r\nn\\
&+&12i{A_3}\sin
\theta \cos \theta ^{2
}{a}^{2}{\beta_1}{\omega}^{2}{r}^{2}+6i{A_3}K\sin 
\theta \cos \theta ^{2}{a}^{2}
{\beta_1}\omega+2i{A_3}K\sin \theta \cos
\theta {Q}^{2}a{\beta_1}\nn\\
&-&2i{A_3} D
K\sin \theta \cos \theta Qa-2
i{A_3}K\sin \theta \cos \theta Qa{
\alpha_1}+2i{A_3}K\sin \theta \cos 
\theta Qa{\alpha_2}+4{A_1} D K\sin
\theta \cos \theta Qar\nn\\
&+&4{A_1}K
\sin \theta \cos \theta Qa{\alpha_1}
r-4{A_1}K\sin \theta \cos \theta 
Qa{\alpha_2}r-12{A_1}K\sin \theta 
\cos \theta ^{2}{a}^{2}{\beta_1}\omega r\nn\\
&-&
4{ A_1}K\sin \theta \cos \theta {Q
}^{2}a{\beta_1}r-3{A_2}K\sin \theta \cos
\theta a{\beta_1}\omega{r}^{2}+4i{A_3}Qa{
\beta_1}\omega{r}^{2}+2i{A_2}KQa{\beta_1}r\nn\\
&+&3{A_2
}K\sin \theta \cos \theta 
^{3}{a}^{3}{\beta_1}\omega-{A_2}K\sin \theta
\cos \theta ^{2}Q{a}^{2}{\alpha_1}+{A_2}K\sin \theta \cos 
\theta ^{2}{Q}^{2}{a}^{2}{\beta_1}\nn\\
&+&{A_2}
D K\sin \theta \cos 
\theta ^{2}Q{a}^{2}
+{ A_2}K\sin \theta
\cos \theta ^{2}Q{a}^{2}{\alpha_2}+2i{A_3}KQa{\beta_1}-{A_2}K \cos 
\theta ^{3}Q{a}^{2}{\beta_1}\nn\\
&-&{A_2}K\sin
\theta {Q}^{2}{\beta_1}{r}^{2}-{A_2} D
K\sin \theta Q{r}^{2}-{A_2}K\sin 
\theta Q{\alpha_2}{r}^{2}+2{ A_2}K\cos \theta
Q{a}^{2}{\beta_1}-{A_2}K\cos \theta Q{
\beta_1}{r}^{2}\nn\\
&+&{ A_2}K\sin \theta Q{\alpha_1
}{r}^{2} \}.\label{lastf}
\eea
We also find similar expansions for all the other components of the gravitational field that are very lengthy and so we do not present them. We note that all functions $A_1,A_2$ and $A_3$ that appear in (\ref{firstf})-(\ref{lastf}) have a linear dependence on $\Delta$ while the other functions such as $\beta_1, \alpha_1, \alpha_2$ do not depend on $\Delta$. A lengthy analysis shows that each component of $f^{\mu\nu}_{ij}$ (that appear in (\ref{compo})) has a structure as 
\be \frac{{\cal A}+{\cal C}\Delta+{\cal E}\Delta^2}{{\cal B}+{\cal F}\Delta+{\cal G}\Delta^2}\label{gens},\ee where the functions ${\cal A},{\cal B},{\cal C},{\cal E},{\cal F},{\cal G}$ do not depend on $\Delta$. Moreover, we notice that the presence of $\rho$ in all components of the graviton $h^{\mu\nu}$, makes the results (\ref{compo}) completely non-separable in terms of coordinates $r$ and $\theta$. However, in calculation of the boundary action, we consider a spherical boundary with radius $r_B$ as the boundary of the near-NHEK geometry. We set the radial coordinate in $\rho$ equal to $r_B$ and so $\rho_B=\rho(r=r_B)$ depends only on $\theta$. As a result, we check out explicitly that any single term in the expansion (\ref{hps}) is completely separable in terms of coordinates $r$ and $\theta$. In other words, we can re-write (\ref{hps}) as
\be
h^{\mu \nu}=\sum _{i,j=+,-} \chi_i\sum_{\aleph}f^{\overline{\mu\nu}}_{1ij\aleph}(r)f^{\overline{\mu\nu}}_{2ij\aleph}(\theta)R_{i}(r)S_j(\theta),\label{hps2}
\ee
where $\chi_+=\Delta_B,\, \chi_-=1$, the over-line means that there is no summation over superscripts $\mu$ and $\nu$. The summation index $\aleph$ in (\ref{hps2}) shows all the individual terms that appear in expansion (\ref{compo}), for example,
\bea
f^{{rr}}_{1++1}(r)&=&
{24{\sqrt{2}}a 
QM\omega r^2},
\\
f^{{rr}}_{1++2}(r)&=&6\sqrt 2 a^2\omega Q r^3,\\
\eea
and
\bea
f^{{rr}}_{2++1}(\theta)&=&
\frac{-\sin ^3\theta\cos^2\theta}{\rho_B^4(3a^2\omega^2\sin^2\theta\cos^2\theta+Qa\omega\sin\theta(1-3\cos^2\theta)+Q^4\sin^2\theta-Q^2)},\,\,\,\,\,\,\,\,\,\,\\
f^{{rr}}_{2++2}(\theta)&=&\frac{-\sin ^3\theta\cos\theta}{\rho_B^4(3a^2\omega^2\sin^2\theta\cos^2\theta+Qa\omega\sin\theta(1-3\cos^2\theta)+Q^4\sin^2\theta-Q^2)}.
\eea

Moreover as we consider the limit $r_B \rightarrow r_+$, 
the metric function $\Delta_B=\Delta(r=r_B)$ approaches to zero on the boundary. The smallness of $\Delta_B$ enables us to find out the most leading divergent term of any expression like (\ref{compo}) on the boundary, that contains a combination of powers of $\Delta$ and general structure (\ref{gens}). 
We notice from (\ref{hps2}) that all different components of graviton in the Kerr background are bi-linear combinations of the radial and angular Teukolsky functions $R_{\pm}=R_{\pm 2}(r)$ and $S_{\pm}(\theta)$. The different solutions to the Teukolsky radial functions $R_{\pm 2}$ were obtained in \cite{Hartman:2009nz} for near region, far region and the matching region where $r << 2r_+$, $r >> r_+(1+\tau_H)$ and $ r_+(1+\tau_H) << r << 2r_+$, respectively. The quantity $\tau_H$ is related to the Hawking temperature by $\tau_H=8\pi M T_H$ and is a very small number for the near extremal Kerr black hole. 
In particular the solutions for the matching region $ r_+(1+\tau_H) << r << 2r_+$ are given by
\bea
R_{+2}&=&A_+x^{\beta-{5}/{2}}+B_+x^{-\beta-\frac{5}{2}}\label{trp},\\
R_{-2}&=&A_-x^{\beta+\frac{3}{2}}+B_-x^{-\beta-\frac{3}{2}}\label{trm},
\eea
where the new coordinate $x$ is given by $x=\frac{r-r_+}{r_+}$ and 
\bea
A_{\pm}&=&\frac{{\Gamma \left( {2\beta } \right)\Gamma \left( {1 \mp 2- i{n}} \right)}}{{\Gamma \left( {\frac{1}{2} + \beta - i\left( {{n} - m} \right)} \right)\Gamma \left( {\frac{1}{2} + \beta \mp 2 - im} \right)}}\tau_H^{-\beta+\frac{1}{2}-i\frac{n}{2}}=\mathcal{A}_ \pm\tau_H^{-\beta+\frac{1}{2}-i\frac{n}{2}}\label{Stro1},\\
{B}_\pm & =& \frac{{\Gamma \left( { - 2\beta } \right)\Gamma \left( {1 \mp 2 - i{n}} \right)}}{{\Gamma \left( {\frac{1}{2} - \beta - i\left( {{n} - m} \right)} \right)\Gamma \left( {\frac{1}{2} - \beta \mp 2 - im} \right)}}
\tau_H^{\beta+\frac{1}{2}-i\frac{n}{2}}=\mathcal{B}_ \pm\tau_H^{\beta+\frac{1}{2}-i\frac{n}{2}}\label{Stro2}.
\eea
In (\ref{Stro1}) and (\ref{Stro2}), $m$ is the azimuthal number, ${n} = \frac{4(\omega - m\Omega_H)M}{\tau_H}$ where $\Omega_H=\frac{a}{r_+^2+a^2}$ shows the angular velocity of the horizon. The constant $\beta$ is 
\begin{equation}
\beta ^2 = \frac{1}{4} + K_l ^{(2)} - 2m^2\label{beta1},
\end{equation}
where $K_l ^{(2)}$ is related to the separation constant $\bar\lambda$ by $K_l ^{(2)} =\bar \lambda+2am\omega+2$ and we choose $K_l ^{(2)} \geq 2m^2-\frac{1}{4}$ to have a real $\beta$.
As we mentioned before, for the near extremal Kerr black hole $\tau_H$ is a very small number and so for a fixed $n$, 
we find $\omega \sim m\Omega_H$. So, we only consider the graviton modes that their energies are almost equal to the super-radiant bound in the background of a near extremal black hole. Moreover, for a near extremal Kerr black hole, $\Omega_H\sim\frac{1}{2a}$ and so we find a relation between the energy and azimuthal number as $2a\omega \sim m$.
In near horizon where $x \rightarrow 0$, we rewrite the solutions (\ref{trp}) and (\ref{trm}) as
\begin{align}
R_ + = \frac{N_+}{\tau _H^{2 + in/2}} \left( {\mathcal{A}_ + \left( {\frac{r}{{\tau _H }}} \right)^{\beta - 5/2} + \mathcal{B}_ + \left( {\frac{r}{{\tau _H }}} \right)^{ - \beta - 5/2} } \right) + ...,\label{Rpsol}
\end{align}
\begin{align}
R_ - = N_- \tau _H^{2 - in/2} \left( {\mathcal{A}_ - \left( {\frac{r}{{\tau _H }}} \right)^{\beta + 3/2} + \mathcal{B}_ - \left( {\frac{r}{{\tau _H }}} \right)^{ - \beta + 3/2} } \right) + ....\label{Rmsol}
\end{align}
In the near horizon limit, the Teukolsky radial functions satisy 
\be
{\cal D}_0^4R_{-2}={\cal W}R_{+2}\label{NOR},
\ee
where the Starobinsky constant ${\cal W}$ is equal to
\be
\vert {\cal W} \vert ^2=\bar \lambda (\bar \lambda +2)^2-8\omega^2\bar \lambda \{\alpha^2(5\bar \lambda+6)-12a^2)+144\omega ^4(\alpha^4+M^2).
\ee
Using equation (\ref{NOR}), we find the ratio of the coefficients 
$N_+$ and $N_-$ in the near horizon limit, is given by
\begin{equation}
\frac{{N_ - }}{{N_ + }} = \frac{\{\bar \lambda^4+4\bar \lambda^3+(4+10m^2)\bar \lambda^2+36m^2\bar \lambda+9m^2(m^2+4)\}^{1/2}r_+^4}{n(n+2i)(n^2+1)} 
\label{relativeNorm}.
\end{equation}

\section{The boundary action for the gravitational perturbation}
\label{sec-bacalc}

In this section, we calculate the on-shell boundary action (\ref{boundact}) and then find the two-point function for the conformal energy-momrntum tensor operators. We find the following relation for the measure function of the boundary action as
\be
\sqrt{-g}=\sqrt{-g^{(0)}}\{1+\frac{\epsilon}{2}\tilde h+\frac{\epsilon^2}{8}(\tilde h^2-h_{\mu\nu}\tilde h^{\mu\nu})+{ O}(\epsilon^3)\}\label{detgexp},
\ee
where $g=\det(g_{\mu\nu})$, $g^{(0)}=\det(g^{(0)}_{\mu\nu})$, $\tilde h^{\mu\nu}=g^{(0)\mu\rho}g^{(0)\nu\sigma}\tilde h_{\rho\sigma}$ and $\tilde h=g^{(0)}_{\alpha\beta}\tilde h^{\alpha\beta}$. Moreover, we find that the Christoffel affine connection can be expanded in terms of $\epsilon$ as
\be
\Gamma^\rho_{\mu\nu}=\Gamma^{(0)\rho}_{\mu\nu}+\epsilon\Gamma^{(1)\rho}_{\mu\nu}+\epsilon^2\Gamma^{(2)\rho}_{\mu\nu}
+O(\epsilon^3)\label{Gammaexp},
\ee
where
\bea
\Gamma^{(0)\rho}_{\mu\nu}&=&\frac{1}{2}g^{(0)\rho\lambda}\{\partial _\mu g^{(0)}_{\lambda\nu}
+\partial _\nu g^{(0)}_{\mu\lambda}
-\partial _\lambda g^{(0)}_{\mu\nu}
\},\\
\Gamma^{(1)\rho}_{\mu\nu}&=&\frac{1}{2}\{g^{(0)\rho\lambda}(\partial _\mu \tilde h_{\lambda\nu}
+\partial _\nu \tilde h_{\mu\lambda}
-\partial _\lambda \tilde h_{\mu\nu})-\tilde h^{\rho\lambda}(\partial _\mu g^{(0)}_{\lambda\nu}
+\partial _\nu g^{(0)}_{\mu\lambda}
-\partial _\lambda g^{(0)}_{\mu\nu})\},\\
\Gamma^{(2)\rho}_{\mu\nu}&=&-\frac{1}{2}\tilde h^{\rho\lambda}\{\partial _\mu \tilde h_{\lambda\nu}
+\partial _\nu \tilde h_{\mu\lambda}
-\partial _\lambda \tilde h_{\mu\nu}\}.
\eea
Substituting expressions (\ref{detgexp}) and (\ref{Gammaexp}) in (\ref{boundact}), we find the only relevant terms in the boundary action that lead to the proper two-point function are
\be
S^{(2)}_B=\int_{r=r_B} d^3x \sqrt{-g^{(0)}} \{g^{(0)r\nu}\Gamma^{(2)\sigma}_{\nu\sigma}+\Gamma^{(1)\sigma}_{\nu\sigma}(\frac{1}{2}g^{(0)r\nu}\tilde h-\tilde h^{r\nu})+\Gamma^{(0)\sigma}_{\nu\sigma}(-\frac{1}{2}\tilde h\tilde h^{r\nu}+\frac{g^{(0)r\nu}}{8}(\tilde h^2-\tilde h_{\rho\lambda}\tilde h^{\rho\lambda}))\}.\label{bounon}
\ee
We notice the different components of the affine connections that appear in the integrand of $S^{(2)}_B$ are
\be
\Gamma^{(0)\sigma}_{r\sigma}=\frac{1}{2}g^{(0)\sigma\lambda}g^{(0)}_{\lambda\sigma,r},
\ee
\be
\Gamma^{(1)\sigma}_{r\sigma}=\frac{1}{2}\{g^{(0)\sigma\lambda}(\partial _r \tilde h_{\lambda\sigma}
+\partial _\sigma \tilde h_{r\lambda}
-\partial _\lambda \tilde h_{r\sigma})-\tilde h^{\sigma\lambda}g^{(0)}_{\lambda\sigma,r}
\},
\ee
and
\be
\Gamma^{(2)\sigma}_{r\sigma}=-\frac{1}{2}\tilde h^{\sigma\lambda}\partial_r\tilde h_{\lambda\sigma}.
\ee
We write the gravitational perturbation $\tilde h_{\alpha\beta}$ as $\frac{1}{2}( h_{\alpha\beta}e^{-i\omega t+im\phi}+h^*_{\alpha\beta}e^{i\omega t-im\phi})$, and we find that the boundary action (\ref{bounon}) can be written as 
\be
S^{(2)}_B=\int_{r=r_B} d^3x \sqrt{-g^{}}\{h^{*\alpha\beta}\Psi_{\alpha\beta\gamma\delta}h^{\gamma\delta}+c.c.\},
\label{bound3}\ee
where
\bea
\Psi_{\alpha\beta\gamma\delta}&=&-\frac{g^{rr}}{8}\{\partial_r(g_{\alpha\gamma}g_{\beta\delta})+g_{\alpha\gamma}g_{\beta\delta}\partial _r
\}
+\frac{g^{rr}}{16}g_{\alpha\beta}\{g_{\gamma\delta}\partial_r-g_{\delta\gamma,r}\}-\frac{1}{8}g_{\alpha}^rg_{\beta}^r\{g_{\gamma\delta}\partial_r-g_{\delta\gamma,r}\}\nn\\
&&-\frac{1}{16}g^{\sigma\lambda}g_{\sigma\lambda,r}g_{\alpha\beta}g_\gamma^rg_\delta^r+\frac{g^{rr}}{64}
g^{\sigma\lambda}g_{\sigma\lambda,r}(g_{\alpha\beta}g_{\gamma\delta}-g_{\alpha\delta}g_{\beta\delta})\label{integrand}.
\eea
We note that in equations (\ref{bound3}) and (\ref{integrand}), for simplicity in notation, we omit the superscript $^{(0)}$ for the Kerr background.
From equation (\ref{hps2}) on the boundary $r=r_B$, we find
\bea
h^{\mu \nu}(r_B)&=&\sum _{\aleph,j=+,-} \Delta_Bf^{\overline{\mu\nu}}_{1+j\aleph}(r_B)f^{\overline{\mu\nu}}_{2+j\aleph}(\theta)R_{+}(r_B)S_j(\theta)+\sum _{\aleph,j=+,-} f^{\overline{\mu\nu}}_{1-j\aleph}(r_B)f^{\overline{\mu\nu}}_{2-j\aleph}(\theta)R_{-}(r_B)S_j(\theta)\nn\\
&\equiv&h^{\mu \nu}_{B+}+h^{\mu \nu}_{B-},
\label{hps3}
\eea
where we define the boundary quantities
\be
\frac{h^{\mu \nu}_{B+}}{R_{B+}}=\sum _{\aleph,j=+,-} \Delta_Bf^{\overline{\mu\nu}}_{1+j\aleph}(r_B)f^{\overline{\mu\nu}}_{2+j\aleph}(\theta)S_j(\theta)\label{hBplus},
\ee
and
\be
\frac{h^{\mu \nu}_{B-}}{R_{B-}}=\sum _{\aleph,j=+,-} f^{\overline{\mu\nu}}_{1-j\aleph}(r_B)f^{\overline{\mu\nu}}_{2-j\aleph}(\theta)S_j(\theta)\label{hBminus}.
\ee
In equations (\ref{hBplus}) and (\ref{hBminus}), $R_{B\pm}$ means $R_{\pm}(r_B)$.
We find that the gravitational perturbation field (\ref{hps2}) 
is given by
\be
h^{\mu \nu}=\frac{h^{\mu \nu}_{B+}}{R_{B+}}R_+(r)+\frac{h^{\mu \nu}_{B-}}{R_{B-}}R_-(r),\label{hhh}
\ee
in terms of boundary fields (\ref{hBplus}) and (\ref{hBminus}). In other words, the gravitational perturbation field can be split into two terms that in each term the non-radial dependence is included in the boundary fields $h^{\mu \nu}_{B\pm}$. Moreover, we notice that (\ref{hhh}) can be written as 
\be
h^{\mu\nu}=
\frac{h_{B+}^{\mu\nu}}{R_{B+}}
R_+(r)+
\frac{1}{\Delta_B}F_{++}^{-+}(\theta)R_-(r)(\frac{h_{B+}^{\mu\nu}}{R_{B+}})_++
\frac{1}{\Delta_B}F_{+-}^{--}(\theta)R_-(r)(\frac{h_{B+}^{\mu\nu}}{R_{B+}})_-,\label{hmunuB}
\ee
where $(\frac{h_{B+}^{\mu\nu}}{R_{B+}})_{\pm}$ refers to the two terms in equation (\ref{hBplus}) with $j=\pm$ for the boundary field $\frac{h_{B+}^{\mu\nu}}{R_{B+}}$. The $\theta$-depedent functions $F_{++}^{-+}$ and $F_{+-}^{--}$ are given by
\be
F_{++}^{-+}(\theta)=\frac{\sum _{\aleph}f^{\overline{\mu\nu}}_{1-+\aleph}(r_B)f^{\overline{\mu\nu}}_{2-+\aleph}(\theta)}{\sum _{\aleph}f^{\overline{\mu\nu}}_{1++\aleph}(r_B)f^{\overline{\mu\nu}}_{2++\aleph}(\theta)},
\ee
and
\be
F_{+-}^{--}(\theta)=\frac{\sum _{\aleph}f^{\overline{\mu\nu}}_{1--\aleph}(r_B)f^{\overline{\mu\nu}}_{2--\aleph}(\theta)}{\sum _{\aleph}f^{\overline{\mu\nu}}_{1+-\aleph}(r_B)f^{\overline{\mu\nu}}_{2+-\aleph}(\theta)},
\ee
respectively. After a lengthy calculation, we find that the second and third terms in (\ref{hmunuB}) are the dominant terms in the boundary action (\ref{bound3}), which is 
\bea
S^{(2)}_B&=&\frac{1}{\Delta_B^2}\int_{r=r_B} d^3x \sqrt{-g^{}}\{[F_{++}^{-+*}(\theta)(\frac{h_{B+}^{\alpha\beta*}}{R_{B+}^*})_++
F_{+-}^{--*}(\theta)(\frac{h_{B+}^{\alpha\beta*}}{R_{B+}^*})_-][F_{++}^{-+}(\theta)(\frac{h_{B+}^{\gamma\delta}}
{R_{B+}})_++
F_{+-}^{--}(\theta)(\frac{h_{B+}^{\gamma\delta}}{R_{B+}})_-]\nn\\
&&R_-^*(r)\Psi_{\alpha\beta\gamma\delta}R_-(r)+c.c.\}.\label{sss}
\eea
To calculate the integrals in (\ref{sss}), we need the derivatives of the radial Teukolsky functions (\ref{Rpsol}), (\ref{Rmsol}) in near horizon limit where $r_B\rightarrow r_+$. We find that the derivatives of radial Teukolsky functions are
\bea
\partial _rR_+(r)&=&(\beta-\frac{5}{2})\frac{1}{r}R_+(r)-2\beta {\cal B}_+N_+\tau_H^{-in/2+\beta+1/2}r^{-\beta-7/2},\label{drp}\\
\partial _rR_-(r)&=&(\beta+\frac{3}{2})\frac{1}{r}R_-(r)-2\beta {\cal B}_-N_-\tau_H^{-in/2+\beta+1/2}r^{-\beta+1/2}\label{drm}.
\eea
We note that the second term in right hand side of equation (\ref{drp}) (or (\ref{drm}) is of the order of $\tau_H^{\beta+1/2}$. For near extremal Kerr black hole $\tau_H \rightarrow 0$ and so we
can ignore the second term in right side of equation (\ref{drp}) (and (\ref{drm})), compared to the first term where $\tau_H \rightarrow 0$. 
We show the $(t,\phi)$-dependence of the boundary fields by ${\cal H}_{B+}^{{\alpha\beta}}(t,\phi)$ and so rewrite the field $h^{\alpha\beta}_{B+}$ as 
\be
h^{\alpha\beta}_{B+}(t,\theta,\phi)={\cal H}_{B+}^{\overline{\alpha\beta}}(t,\phi)\Theta^{\overline{\alpha\beta}}_{\pm}(\theta),
\ee
where $
{\cal H}_{B+}^{\overline{\alpha\beta}}(t,\phi) $ is equal to $\Delta_Be^{-i\omega t+im\phi}R_{B+}$ for $\alpha,\beta=t,r,\theta,\phi$
and 
\be
\Theta^{\overline{\alpha\beta}}_\pm=\sum_{\aleph} [f^{\overline{\alpha\beta}}_{1+{\pm}\aleph}(r_B)f^{\overline{\alpha\beta}}_{2+\pm\aleph}(\theta)] S_{\pm}(\theta).
\ee
As a result, the leading term in the boundary action becomes
\bea
S_B^{(2)}&=&\frac{1}{\Delta_B^2}\int_{r=r_B} d^3x \sqrt{-g^{}}\{[F_{++}^{-+*}(\theta){\Theta_{+}^{\overline{\alpha\beta}*}}+
F_{+-}^{--*}(\theta)\Theta_{-}^{\overline{\alpha\beta}*}](\Psi_{B})_{\alpha\beta\gamma\delta}(r_B,\theta)[F_{++}^{-+}(\theta)\Theta_{+}^{\overline{\gamma\delta}}+
F_{+-}^{--}(\theta)\Theta_{-}^{\overline{\gamma\delta}}]\nn\\
&\times&\frac{R_-^*(r_B)R_-(r_B)}{{R_{B+}^*{R_{B+}}}}{\cal H}_{B+}^{\overline{\alpha\beta}*}(t,\phi){\cal H}_{B+}^{\overline{\gamma\delta}}(t,\phi)+c.c.\}\label{action11},
\eea
where
\bea
(\Psi_B)_{\alpha\beta\gamma\delta}(r_B,\theta)&=&\{-\frac{g^{rr}}{8}\partial_r(g_{\alpha\gamma}g_{\beta\delta})
+\frac{g^{rr}(\beta+3/2)}{8r_B}(\frac{1}{2}g_{\alpha\beta}g_{\gamma\delta}-g_{\alpha\gamma}g_{\beta\delta}-\frac{g^r_\alpha g^r_\beta g_{\gamma\delta}}{g^{rr}})
\nn\\
&+&\frac{1}{8}(g_{\alpha}^rg_{\beta}^r-\frac{1}{2}g^{rr}g_{\alpha\beta})g_{\delta\gamma,r}-\frac{1}{16}g^{\sigma\lambda}g_{\sigma\lambda,r}g_{\alpha\beta}g_\gamma^rg_\delta^r+\frac{g^{rr}}{64}
g^{\sigma\lambda}g_{\sigma\lambda,r}(g_{\alpha\beta}g_{\gamma\delta}-g_{\alpha\delta}g_{\beta\delta})\}\vert_{r=r_B}.\nn\\
&&\label{integrand2}
\eea
Using equations (\ref{Rpsol}) and (\ref{Rmsol}) for the Teukolsky radial functions, we find that the ratio of boundary Teukolsky functions $\frac{R_-^*(r_B)R_-(r_B)}{{R_{B+}^*{R_{B+}}}}$ in equation (\ref{action11}), is given by
\be
\frac{R_-^*(r_B)R_-(r_B)}{{R_{B+}^*{R_{B+}}}}=\frac{{\cal N}r_B^8\{\vert {\cal A}_- \vert ^2+( {\cal A}_- {\cal B}_-^*+{\cal A}_-^*{\cal B}_-)(\frac{r_B}{\tau_H})^{-2\beta}+\vert {\cal B}_- \vert ^2(\frac{r_B}{\tau_H})^{-4\beta}\}}{n^2(n^2-4)(n^2+1)^2\vert {\cal A}_+ \vert ^2}\label{ratioR},
\ee
where ${\cal N}=\bar \lambda^4+4\bar \lambda^3+(4+10m^2)\bar \lambda^2+36m^2\bar \lambda+9m^2(m^2+4)$.
We find the two-point function of the boundary energy-momentum tensor operators  ${\cal O_{\alpha\beta}}$ by taking the functional derivative of the action (\ref{action11}) with respect to the rescaled boundary gravitational field $\hat {\cal H}_{B+}^{{\alpha\beta}}$ as $<{\cal O}_{\alpha\beta}{\cal O}_{\gamma\delta}>=\frac{{\delta ^2 S_B^{(2)} }}{{\delta\hat {\cal H}_{B+}^{{{ \alpha\beta}}}\delta\hat{\cal H}_{B+}^{{{\gamma\delta}}} }} 
$. We find
\be
<{\cal O}_{\alpha\beta}{\cal O}_{\gamma\delta}>=\frac{{\delta ^2 S_B^{(2)} }}{{\delta\hat {\cal H}_{B+}^{{{ \alpha\beta}}}\delta\hat{\cal H}_{B+}^{{{\gamma\delta}}} }} 
= r_B^{2\beta-8}\frac{{\left( {R_ - ^B } \right)^* R_ - ^B }}{{\left( {R_ + ^B } \right)^* R_ + ^B }} {\cal{F}}_{\alpha\beta\gamma\delta},\label{2pt1}
\ee
where 
\bea
{\cal{F}}_{\alpha\beta\gamma\delta} &=& \int_0^\pi d\theta \sin(\theta) 
\{[F_{++}^{-+*}(\theta){\Theta_{+}^{\overline{\alpha\beta}*}}+
F_{+-}^{--*}(\theta)\Theta_{-}^{\overline{\alpha\beta}*}](\Psi_{B})_{\alpha\beta\gamma\delta}(r_B,\theta)[F_{++}^{-+}(\theta)\Theta_{+}^{\overline{\gamma\delta}}+
F_{+-}^{--}(\theta)\Theta_{-}^{\overline{\gamma\delta}}]
\nn\\
&+&[F_{++}^{-+}(\theta){\Theta_{+}^{\overline{\alpha\beta}}}+
F_{+-}^{--}(\theta)\Theta_{-}^{\overline{\alpha\beta}}](\Psi_{B})_{\alpha\beta\gamma\delta}(r_B,\theta)[F_{++}^{-+*}(\theta)\Theta_{+}^{\overline{\gamma\delta}*}+
F_{+-}^{--*}(\theta)\Theta_{-}^{\overline{\gamma\delta}*}]
\}\rho_B^2 \label{thetaint}.
\eea
We note that the rescaled boundary gravitational field $\hat {\cal H}_{B+}^{{\alpha\beta}}$is given by $\hat {\cal H}_{B+}^{{\alpha\beta}}=\frac{ {\cal H}_{B+}^{{\alpha\beta}}+ {\cal H}_{B+}^{{\alpha\beta*}}}{2\Delta_B r_B^{\beta-4}}$.
Using (\ref{Stro1}) and (\ref{Stro2}), we can rewrite (\ref{2pt1}) as
\be
<{\cal O}_{\alpha\beta}{\cal O}_{\gamma\delta}>
= M^8r_B^{2\beta-8} {\cal{F}}_{\alpha\beta\gamma\delta}\{1+M^{-2\beta}\frac{\cal N}{\cal C}G_R^*+M^{-2\beta}\frac{{{\cal N}}}{\cal C^*}G_R+{\cal N}M^{-4\beta}\vert G_R\vert^2\}\label{twop},
\ee
where we find that $G_R$ is given by
\be
G_R=T_R^{2\beta}\frac{\Gamma(-2\beta)\Gamma(\beta-3/2-im)\Gamma(1/2+\beta-i(n-m))}{\Gamma(2\beta)\Gamma(5/2-\beta-im)\Gamma(1/2-\beta-i(n-m))},\label{GRRR}
\ee
and
\be
{\cal C}=(\bar \lambda-2m^2-2im\beta)(\bar\lambda-2m^2-2im\beta+2).
\ee

As we notice, 
the second and third terms in bracket in equation (\ref{twop}) are complex conjugate of each other and so we find a real-valued two-point function for the boundary conformal operators. However according to similar analysis in \cite{spin1}, \cite{Becker:2012vda} and \cite{son}, we can drop the second term in (\ref{twop}) to find a complex-valued two-point function for the boundary conformal operators.
We also notice that the coefficients of term $G_R{\cal{F}}_{\alpha\beta\gamma\delta}$ in (\ref{twop}) depend on the momentum and do not contribute to the $G_R$, given by (\ref{GRRR}). Similar momentum-dependent coefficients previously have been found for the two-point functions of spin-1/2 operators \cite{Becker:2012vda} as well as spin-1 operators \cite{spin1}. Moreover, we note that the first term in bracket in equation (\ref{twop}) is just a constant, compared to the third term that is proportional to $G_R$. The last term in (\ref{twop}) is very small compared to the third term, since according to equation (\ref{ratioR}), the ratio of the last term to the third term is proportional to $(\frac{\tau_H}{M})^{2\beta}$ where $\tau_H\rightarrow 0$. Hence we conclude that the field theoretical two-point function for spin-2 conformal operators ${\cal O_{\alpha\beta}}$ on the boundary can be described by
$G_R{\cal{F}}_{\alpha\beta\gamma\delta}$. The retarded Green's functions for fields with different spins were proposed in \cite{Hartman:2009nz} and \cite{Chen:2010ni}. We note that 
$G_R$, as a part of two-point function $G_R{\cal{F}}_{\alpha\beta\gamma\delta}$ for the boundary conformal operators, is in perfect agreement with the proposed retarded Green's function for spin-2 fields. We also find that the absorption cross section of fields (with spin 2) is in exact agreement with the result of paper \cite{Bredberg:2009pv} if we apply the optical theorem to the retarded Green's function (\ref{GRRR}).

\section{Correlation function of spin-2 operators in CFT }
\label{sec-CFT}

The correlation functions for spin-2 operators in a conformal field theory have been found in the context of gravitons in AdS/CFT correspondence \cite{spin2cft,spin2cft2}. The two-point correlation function for the spin-2 conformal operators on the boundary of a $d+1$-dimensional AdS is given by
\be
\left\langle{\cal T}_{\alpha \beta}(\vec x){\cal T}_{\gamma \rho}(\vec y)\right\rangle=\frac{\cal C}{\vert \vec x-\vec y \vert ^{2d}}\{\frac{1}{2}U_{\alpha\gamma}(\vec x-\vec y)U_{\beta\rho}(\vec x-\vec y)+U_{\alpha\rho}(\vec x-\vec y)U_{\beta\gamma}(\vec x-\vec y)-\frac{\delta_{\alpha\beta}\delta_{\gamma\rho}}{d}\}\label{TT},
\ee
where ${\cal C}$ is a constant that depends on $d$ and$\vec x$ denotes a point on the boundary of AdS and $U_{\alpha\beta}(\vec x)=\delta_{\alpha\beta}-\frac{2x_\alpha x_\beta}{x^2}$. The result is based on using the radiation gauge and time slicing formulation.

We know that using other gauge conditions, the two-point function (\ref{TT}) takes a more general form such as
\be
\left\langle {\cal T}_{\alpha \beta}(\vec x){\cal T}_{\gamma \rho}(\vec y) \right\rangle = \frac{V_{\alpha\beta\gamma\rho}(\vec x-\vec y)}
{{\left| {x - y} \right|^{2d } }}
\label{OOCFT},
\ee
where the functional form of tensor $V_{\alpha\beta\gamma\rho}$
depends on the gauge condition. As the near horizon of near extremal Kerr black hole contains a copy of 2-dimensional AdS, we expect that the two-point function of spin-2 conformal operators is given by (\ref{OOCFT}). However, we notice that it would be a very difficult and even impossible task to find the functional form of tensor $V_{\alpha\beta\gamma\rho}$ for the gauge conditions $\Psi_1^{(1)}=\Psi_3^{(1)}=0$. We noticed that these proper gauge conditions enable us to write the spin coefficients of the Kerr black hole, directly in terms of the radial and angular Teukolsky functions. However, comparing the factorized two-point functions of the spin-2 conformal operators (\ref{OOCFT}) with the field-theoretic result $G_R{\cal F}_{\alpha\beta\gamma\delta}$ suggests that we may associate the retarded Green's function $G_R$ to ${\vert \vec x-\vec y \vert^{-2d}}$. Of course, we associate the remaining part of the two-point function ${\cal F}_{\alpha\beta\gamma\delta}$ with the Lorentz indices, to the tensor $V_{\alpha\beta\gamma\delta}$ in (\ref{OOCFT}). We also note that we can find the finite temperature two-point function of scalars \cite{Hartman:2009nz,Chen:2010ni}, spin-$1/2$ fermions \cite{Becker:2012vda} as well as spin-1 fields \cite{spin1} from the expression 
\be
\left\langle {\mathcal{O}{\mathcal{O}}} \right\rangle \sim \left( {\frac{{\pi T_R }}
{{\sinh \left( {\pi T_R t_{12}^+ } \right)}}} \right)^{2h_R } \left( {\frac{{\pi T_L }}
{{\sinh \left( {\pi T_L t_{12}^- } \right)}}} \right)^{2h_L } \label{OOtorus},
\ee
by proper identification of left and right conformal weights $h_L$ and $h_R$. In (\ref{OOtorus}),
$1/T_L$ and $1/T_R$ are the circumferences of two fundamental circles of a torus. 
In fact, for scalars the conformal weights are equal 
$h_R =h_L=\beta + 1/2$. For spin-$1/2$ operators, the conformal weights are
$h_R=\beta+1/2$ and $h_L =\beta$ and for spin-1 operators, they are given by
$h_R = \beta + 1/2$ and $h_L = \beta - 1/2$. 
The Fourier transform of the two-point function (\ref{OOtorus}) is given by
\be
{\cal F}({\left\langle {\mathcal{O}{\mathcal{O}}} \right\rangle})\sim {T_R}^{2\beta} \frac{{\Gamma \left( {1 - 2h_R } \right)\Gamma \left( {1 - 2h_L } \right)}}{{\Gamma \left( {1 - h_R -in_R } \right)\Gamma \left( {1 - h_R +in_R } \right)\Gamma \left( {1 - h_L -in_L} \right)\Gamma \left( {1 - h_L +in_L} \right)}}\label{JJ},
\ee
where $n_R$ and $n_L$ stand for $\frac{i\omega_R}{2\pi T_R}$ and $\frac{i\omega_L}{2\pi T_L}$ respectively. If we consider the right and left conformal weights $h_R = \beta + 1/2$ and $h_L = \beta - 3/2$ for spin-2 operators, we then find
\be
{\cal F}({\left\langle {\mathcal{O}{\mathcal{O}}} \right\rangle})\sim {T_R}^{2\beta} \frac{{\Gamma \left( {- 2\beta} \right)\Gamma \left( {1/2+\beta-in_R } \right)\Gamma \left( {\beta-3/2-in_L } \right)}}{{\Gamma \left( {1/2-\beta-in_R} \right)\Gamma \left( {-\beta+5/2-in_L} \right)\Gamma \left( {2\beta} \right)}}\label{JJ2}.
\ee
Moreover, we consider $n_L=m$ and $n_R=n-m$ in (\ref{JJ2}) and find an exact agreement with the retarded Green's function (\ref{GRRR}) that we found as a part of two-point function for spin-2 operators.

\section{Concluding remarks}
\label{sec-conclu}

The energy-momentum tensor operators of the CFT that is dual to the near extremal Kerr black hole couple to the gravitational boundary fields with spin-2 on the near horizon of near extremal Kerr black hole. Hence we can find the two-point function of the energy-momentum tensor operators by finding the explicit form of a boundary action that is bilinear in the boundary fields. In this article, we construct explicitly such a boundary action and vary the action with respect to the boundary fields. We derive the explicit expression for the two-point function and note that the two-point function factorizes into two terms. We show that the result for the two-point function is in agreement with correlation functions of spin-2 conformal operators in a dual CFT. We also note that one part of the result heavily depends on the gauge choice that we made on fixing some of the Weyl scalars. Investigation about the gauge dependence of the two-point function would be an interesting project. Moreover, deriving the two-point function of the energy-momentum tensor operator with the primary fields of the dual CFT would be useful to establish a stronger field-theoretic approach to Kerr/CFT correspondence.

\bigskip
\bigskip

{\Large Acknowledgments}\newline
This work was supported by the Natural Sciences and Engineering Research Council of Canada. This research was supported in part by Perimeter Institute for Theoretical Physics. Research at Perimeter Institute is supported by the Government of Canada through Industry Canada and by the Province of Ontario through Ministry of Economic Development \& Innovation. \newline

\appendix

\section{The spin coefficients and the Weyl scalars}
\label{Ap1}

The Ricci rotation coefficients are defined by
\be
\gamma _{a b c}=e^\mu _a e^\nu _c e_{b\mu;\nu},
\ee
in terms of the vierbein field $e^\mu _a$ and its covariant derivatives. The lower case Latin indices are the tetrad indices $a,b,c=1,\cdots , 4$ while the Greek indices show the space-time indices $\mu,\nu=1, \cdots , 4$.
The spin coefficients are different combinations of the Ricci rotation coefficients. They are explicitly given by
\be
\kappa = \gamma _{311}, \sigma=\gamma_{313}, \lambda=\gamma_{244}, \nu=\gamma_{242},\rho=\gamma_{314},\mu=\gamma_{243},\tau=\gamma_{312},\pi=\gamma_{241},
\ee
\be
\alpha=\frac{\gamma_{214}+\gamma_{344}}{2}, \beta=\frac{\gamma_{213}+\gamma_{343}}{2},
\epsilon=\frac{\gamma_{211}+\gamma_{341}}{2},\gamma=\frac{\gamma_{212}+\gamma_{342}}{2}.
\ee

Moreover, the Weyl tensor $C_{abcd}$ is defined by
\be
C_{abcd}=R_{abcd}+\frac{1}{2}\{\eta_{bc}R_{ad}+\eta_{ad}R_{bc}-\eta_{ac}R_{bd}-\eta_{bd}R_{ac}+\frac{\eta_{ac}\eta_{bd}-\eta_{ad}\eta_{bc}}{3}R\},
\ee
in terms of the Riemann tensor $R_{abcd}$ and the Ricci tensor $R_{ab}$. The normalization matrix $\eta_{ab}$ is defined by $l_a\cdot l_b=\eta_{ab}$. The Weyl scalars $\Psi_{0},\cdots, \Psi_4$ represent the independent components of the Weyl tensor which are given by different contractions of the Weyl tensor with the Newman-Penrose null vectors. They are given by $\Psi_0=-C_{1313},\Psi_1=-C_{1213},\Psi_2=-C_{1342},\Psi_3=-C_{1242},\Psi_4=-C_{2424}$.

\end{document}